\documentclass[prl,twocolumn,preprintnumbers,amsmath,amssymb]{revtex4}

\usepackage{graphicx}
\usepackage{amssymb,amsmath}
\usepackage{dcolumn}
\usepackage{bm}
\usepackage{color}
\usepackage{enumerate}
\usepackage{epstopdf}
\usepackage{hyperref}

\begin{document}

\title{Anomalous radiation reaction in a circularly polarized field}

\author{O. V. Kibis}\email{oleg.kibis(c)nstu.ru}

\affiliation{Department of Applied and Theoretical Physics,
Novosibirsk State Technical University, Karl Marx Avenue 20,
Novosibirsk 630073, Russia}

\begin{abstract}
Quantum corrections to electron dynamics in a circularly polarized electromagnetic field are found within the Floquet theory of periodically driven quantum systems. It is demonstrated that emission of photons by an electron rotating under the field leads to the quantum recoil force acting on the electron perpendicularly to the velocity of its forward movement, which differs crucially from the known classical recoil force directed oppositely to the velocity. Physically, such an anomalous radiation reaction arises from the one-loop QED correction to the photon emission and has no analog within the classical electrodynamics. Possible manifestations of this macroscopic QED effect are discussed for electrons in strong laser fields.
\end{abstract}

\maketitle

{\it Introduction}---The laws of nature describing the motion of charged particles in electromagnetic fields lie in the basis of most physical processes in the Universe and, therefore, they are permanently in the focus of attention despite their long prehistory.  Since a charged particle driven by a high-frequency field makes oscillating movement, it emits electromagnetic radiation. If the field is strong enough, the radiation reaction dominates the particle motion and needs the quantum electrodynamics (QED) to be described accurately. Although this research area is exciting nowadays due to achievements in the laser generation of strong fields~\cite{Gonoskov_2022,Fedotov_2023,Piazza_2012,Piazza_2008,Piazza_2010_1,Piazza_2010,Piazza_2009,Ilderton_2013,Cole_2018,Poder_2018}, the features of the QED corrections depending on the field polarization still wait for detailed analysis. To fill partially this gap, the emission of photons by an electron rotating under a circularly polarized field was considered within the Floquet theory which is conventionally used to describe various periodically driven quantum systems~\cite{Goldman_2014,Bukov_2015,Eckardt_2015,Casas_2001,Kibis_2022}. Solving this problem, it was found that the photon emission leads to the quantum recoil force acting on the emitting electron perpendicularly to the velocity of its forward movement. This result is surprising since the classical Lorentz-Abraham-Dirac (LAD) force --- which was introduced into physics at the dawn of electrodynamics to describe the radiation reaction and takes deserved place in textbooks~\cite{Landau_2,Barut_book} --- is directed oppositely to the velocity. The present Letter is devoted to the first theoretical analysis of this anomalous radiation reaction which can manifest itself in strong laser fields.

{\it Model}---For definiteness, let us consider a classically strong circularly polarized homogeneous electromagnetic field with the zero scalar potential and the vector potential
\begin{equation}\label{A0L}
\mathbf{A}=(A_x,A_y,A_z)=\left(\frac{cE_0}{\omega}\right)(\cos\omega t,\,\sin\omega t,\,0),
\end{equation}
where $E_0$ is the electric field amplitude, $\omega$ is the field frequency, $c$ is the light speed in vacuum, and the field polarization is assumed to be clockwise. Within the framework of classical electrodynamics, the vector potential \eqref{A0L} describes the rotating electric field, $\mathbf{E}=-(1/c)\partial_t\mathbf{A}$, which induces electron rotation with the velocity $v_0=eE_0/m_e\omega$ along circular trajectory of the radius $r_0=|v_0|/\omega$, where $m_e$ is the electron mass and $e$ is the electron charge~\cite{Landau_2}. Proper generalization of the model field \eqref{A0L} to the physically important case of electromagnetic wave will be done in the following. The non-relativistic quantum dynamics of an electron under the field \eqref{A0L} is defined by the Hamiltonian
$
\hat{\cal H}_0=\left(\hat{\mathbf{p}}-{e}\mathbf{A}/c\right)^2/{2m_e},
$
where $\hat{\mathbf{p}}=(\hat{{p}}_x,\,\hat{{p}}_y,\,\hat{{p}}_z)$
is the momentum operator. The exact eigenfunctions of the Hamiltonian are
\begin{align}\label{psiL}
&\psi_\mathbf{k}=\frac{e^{i\mathbf{k}\mathbf{r}}}{\sqrt{\cal{V}}}\,
e^{-i(\varepsilon_\mathbf{k}
+\varepsilon_0)t/\hbar}e^{i(v_0/\omega)(k_x\sin\omega t-k_y\cos\omega t)},
\end{align}
where $\mathbf{r}=(x,y,z)=(r\sin\theta\cos\varphi,\,r\sin\theta\sin\varphi,\,r\cos\theta)$ is the electron radius vector, $\theta$ and $\varphi$ are the polar and azimuth angle, respectively, $\mathbf{k}=(k_x,\,k_y,\,k_z)=
(k\sin\theta_\mathbf{k}\cos\varphi_\mathbf{k},\,k\sin\theta_\mathbf{k}\sin\varphi_\mathbf{k},\,k\cos\theta_\mathbf{k})$ is the electron wave vector, $\varepsilon_\mathbf{k}=\hbar^2{\mathbf{k}}^2/2m_e$ is the kinetic energy of forward movement of the electron, $\varepsilon_0=m_ev_0^2/2$ is the kinetic energy of electron rotation under the field \eqref{A0L}, and $\cal{V}$ is the normalization volume.
The wave functions \eqref{psiL} can be easily proved by direct substitution into the Schr\"odinger equation, $\hat{\cal H}_0\psi_{\mathbf{k}}=i\hbar\partial_t\psi_{\mathbf{k}}$, and should be treated as the Floquet functions~\cite{Goldman_2014,Bukov_2015,Eckardt_2015,Casas_2001} describing dynamics of an electron strongly coupled to the field \eqref{A0L} (dressed by the field). The averaged electron velocity in the Floquet state \eqref{psiL} reads
\begin{equation}\label{vL}
{\mathbf{v}}=\frac{i}{\hbar}\langle\psi_\mathbf{k}|\hat{\cal H}_0\mathbf{r}-\mathbf{r}\hat{\cal H}_0|\psi_\mathbf{k}\rangle=\mathbf{v_k}+\mathbf{v}_0,
\end{equation}
where $\mathbf{v_k}=\hbar\mathbf{k}/m_e$ and $\mathbf{v}_0=v_0(-\cos\omega t,\,-\sin\omega t,\,0)$ are the velocity of forward and rotational movement of the electron, respectively. As expected, the velocity \eqref{vL} exactly coincides with the classical velocity of an electron under a circularly polarized field~\cite{Landau_2}. Next, let us find corrections to this velocity arisen from the emission of photons by the dressed electron.

The interaction between an electron and a photon in the presence of the background field \eqref{A0L} is described by the Hamiltonian $\hat{\cal H}=\hat{\cal H}_0+\hat{V}^{(\pm\mathbf{q})}$, where $\hat{V}^{(\pm\mathbf{q})}$ is the Hamiltonian of electron-photon interaction, and the signs ``$+$'' and ``$-$'' correspond to the absorption and emission of a photon with the wave vector $\mathbf{q}$, respectively. Since the Floquet functions \eqref{psiL} form the complete orthonormal function system for any time $t$, the solution of the Schr\"odinger problem with the Hamiltonian $\hat{\cal H}$ can be sought as an expansion
$
\psi=\sum_{\mathbf{k}'}c_{\mathbf{k}'}(t)\psi_{\mathbf{k}'}.
$
Substituting this expansion into the Schr\"odinger equation, $\hat{\cal H}\psi=i\hbar\partial_t\psi$, we arrive at the Schr\"odinger equation written in the Floquet representation,
\begin{equation}\label{dckL}
i\hbar\frac{dc_{\mathbf{k}'}(t)}{dt}=\sum_{\mathbf{k}}\langle\psi_{\mathbf{k}'}|\hat{V}^{(\pm\mathbf{q})}|\psi_{\mathbf{k}}\rangle c_{\mathbf{k}}(t),
\end{equation}
which defines the expansion coefficients $c_{\mathbf{k}}(t)$. Within the conventional quantum electrodynamics~\cite{Landau_4}, the matrix element of the electron interaction with a single photon reads
\begin{equation}\label{V0L}
\langle\psi_{\mathbf{k}'}|\hat{V}^{(\pm\mathbf{q})}|\psi_{\mathbf{k}}\rangle=\frac{e}{c}
\int_{\cal{V}}\left(\mathbf{j}_{\mathbf{k}'\mathbf{k}}\cdot\mathbf{A}^{(\pm\mathbf{q})}\right)d^3\mathbf{r},
\end{equation}
where $\mathbf{A}^{(\pm\mathbf{q})}=\mathbf{e}^{(\mathbf{\pm q})}\sqrt{2\pi\hbar c/q\cal{V}}\,e^{\pm i(\mathbf{qr}-cqt)}$ is the photon wave function, $\mathbf{q}=(q_x,\,q_y,\,q_z)=(q\sin\theta_\mathbf{q}\cos\varphi_\mathbf{q},\,q\sin\theta_\mathbf{q}\sin\varphi_\mathbf{q},\,q\cos\theta_\mathbf{q})$ is the photon wave vector, ${\mathbf{e}^{(\mathbf{+q})}}^\ast={\mathbf{e}^{(\mathbf{-q})}}\equiv\mathbf{e}^{(\mathbf{q})}$ is the unit vector of photon polarization,
\begin{align}\label{jkL}
&\mathbf{j}_{\mathbf{k}'\mathbf{k}}=\frac{1}{2m_e}(\psi^\ast_{\mathbf{k}'}\hat{\mathbf{p}}\psi_{\mathbf{k}}-
\psi_{\mathbf{k}}\hat{\mathbf{p}}\psi^\ast_{\mathbf{k}'})-\frac{e}{m_ec}\mathbf{A}
\psi^\ast_{\mathbf{k}'}\psi_{\mathbf{k}}\nonumber\\
&=\frac{e^{i(\mathbf{k}-\mathbf{k}')\mathbf{r}}}{\cal{V}}
\sum_{m=-\infty}^{\infty}\mathbf{j}_{\mathbf{k}'\mathbf{k}}^{(m)}e^{-im\omega t}
\end{align}
is the transition current, and $\mathbf{j}_{\mathbf{k}'\mathbf{k}}^{(m)}$ are its Fourier harmonics. Applying the Jacobi-Anger expansion, $e^{i\xi\sin\omega t}=\sum_{m=-\infty}^{\infty}J_m(\xi)e^{im\omega t}$, to find these harmonics, the matrix element \eqref{V0L} can be rewritten as
\begin{align}\label{VnmL}
&\langle\psi_{\mathbf{k}'}|\hat{V}^{(\pm\mathbf{q})}|\psi_{\mathbf{k}}\rangle=
\frac{(2\pi)^3}{\cal{V}}e^{-i(\varepsilon_\mathbf{k}-\varepsilon_{\mathbf{k}\pm\mathbf{q}}\pm \hbar cq)t/\hbar}
\nonumber\\
&\times\sum_{m=-\infty}^{\infty}
V_{\mathbf{k}\pm\mathbf{q},\,\mathbf{k}}^{(m)} e^{-im\omega t}\delta^3(\mathbf{k}-\mathbf{k}'\pm\mathbf{q}),
\end{align}
where
\begin{align}\label{VmmL}
&V_{\mathbf{k}\pm\mathbf{q},\,\mathbf{k}}^{(m)}=e\sqrt{\frac{2\pi\hbar }{cq\cal{V}}}
\Bigg[\left(\mathbf{v_k}\cdot\mathbf{e}^{(\mathbf{\pm q})}\right)J_{m}(\xi_\mathbf{q}^\pm)
\nonumber\\
&-
v_0\frac{{e}^{(\pm \mathbf{q})}_x- i{e}^{(\pm \mathbf{q})}_y}{2}J_{m+1}(\xi_\mathbf{q}^\pm)e^{i\varphi_\mathbf{q}}\nonumber\\
&\left.-
v_0\frac{{e}^{(\pm \mathbf{q})}_x+i{e}^{(\pm \mathbf{q})}_y}{2}
J_{m-1}(\xi_\mathbf{q}^\pm)e^{-i\varphi_\mathbf{q}}\right]e^{im\varphi_\mathbf{q}},
\end{align}
$J_m(\xi_\mathbf{q}^{\pm})$ is the Bessel function of the first kind, and its argument is $\xi_\mathbf{q}^\pm=\pm(v_0/\omega)q\sin\theta_\mathbf{q}$.

\begin{figure}[!h]
\includegraphics[width=0.8\linewidth]{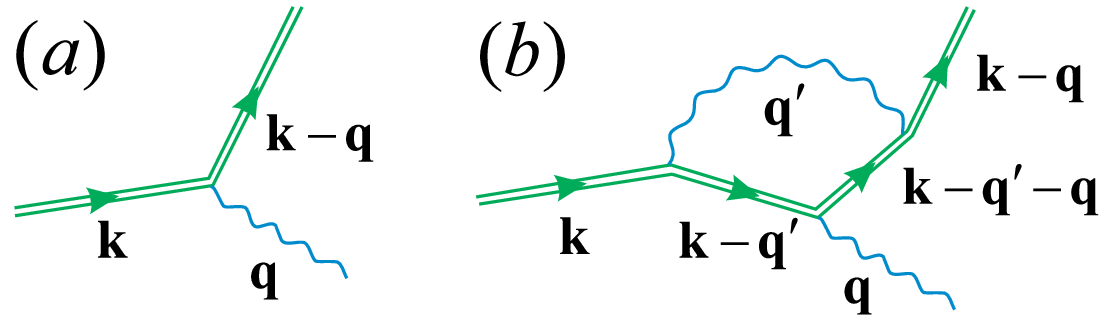}
\caption{Diagram representation of the emission of a photon with the wave vector $\mathbf{q}$ by an electron with the wave vector $\mathbf{k}$ in a background circularly polarized field: (a) the one-vertex emission process; (b) the one-loop emission process through the intermediate state of a virtual photon with the wave vector $\mathbf{q}'$. The double fermion lines indicate the electron strongly coupled to the background field.}
\end{figure}
{\it Results and discussion}---Using the Floquet functions \eqref{psiL} as a basis to expand the electron wave function, we take into account the electron interaction with the strong background field \eqref{A0L} accurately (non-perturbatively), whereas the electron interaction with photons can be considered as a weak perturbation. First of all, let us consider the one-vertex photon emission pictured in Fig.~1a, assuming an electron to be initially in the Floquet state $\psi_\mathbf{k}$. Solving the Schr\"odinger equation \eqref{dckL} within the first order of the conventional perturbation theory~\cite{Landau_3}, we arrive at the amplitude of the one-vertex process,
\begin{equation}\label{ckL}
c^{(a)}_{\mathbf{k}'}(t')=-\frac{i}{\hbar}\int_0^{t'}\langle\psi_{\mathbf{k}'}|\hat{V}^{(-\mathbf{q})}|\psi_{\mathbf{k}}\rangle dt.
\end{equation}
Then the probability of the photon emission per unit time reads
\begin{align}\label{wqL}
&W^{(a)}_\mathbf{k}(\mathbf{q})=\sum_{\mathbf{e}^{(\mathbf{q})}}\sum_{\mathbf{k}'}\frac{d \left|c^{(a)}_{\mathbf{k}'}(t')\right|^2}{dt'}=\frac{2\pi}{\hbar}\sum_{\mathbf{e}^{(\mathbf{q})}}
\left|T^{(a)}_\mathbf{k}(\mathbf{q})\right|^2\nonumber\\
&\times\delta(\varepsilon_{\mathbf{k}-\mathbf{q}}-\varepsilon_{\mathbf{k}}-\hbar\omega+\hbar cq)\approx\frac{\pi^2e^2v_0^2}{cq\cal{V}}
\Big[1+
\cos^2\theta_{\mathbf{q}}\nonumber\\
&+\frac{2v_{\mathbf{k}x}}{c}\cos^2\theta_{\mathbf{q}}\sin\theta_{\mathbf{q}}\cos\varphi_{\mathbf{q}}
+\frac{2v_{\mathbf{k}y}}{c}\cos^2\theta_{\mathbf{q}}\sin\theta_{\mathbf{q}}\sin\varphi_{\mathbf{q}}\nonumber\\
&-\frac{2v_{\mathbf{k}z}}{c}\cos\theta_{\mathbf{q}}\sin^2{\theta_{\mathbf{q}}}\Big]\delta(\varepsilon_{\mathbf{k}-\mathbf{q}}-\varepsilon_{\mathbf{k}}-\hbar\omega+\hbar cq),
\end{align}
where
\begin{align}\label{TaL}
&T^{(a)}_\mathbf{k}(\mathbf{q})=V_{\mathbf{k}-\mathbf{q},\,\mathbf{k}}^{(1)}\approx
-{ev_0}\sqrt{\frac{\pi\hbar }{2cq\cal{V}}}\nonumber\\
&\times\left[
{{e}^{(\mathbf{q})}_x+i{e}^{(\mathbf{q})}_y}+\frac{\left(\mathbf{v_k}\cdot
\mathbf{e}^{(\mathbf{q})}\right)}{c}\left({n}^{(\mathbf{q})}_x+i{n}^{(\mathbf{q})}_y\right)\right]
\end{align}
is the probability amplitude of the one-vertex photon emission, and $\mathbf{n}^{(\mathbf{q})}=\mathbf{q}/q=(\sin\theta_{\mathbf{q}}\cos\varphi_{\mathbf{q}},\,
\sin\theta_{\mathbf{q}}\sin\varphi_{\mathbf{q}},\,\cos\theta_{\mathbf{q}})$
is the unit wave vector~\cite{SM}. In the following, we will assume that the field frequency, $\omega$, is high enough to satisfy the condition $\omega\tau\gg1$,
where
\begin{equation}\label{tauL}
{\tau}=\left[\sum_\mathbf{q}W^{(a)}_\mathbf{k}(\mathbf{q})\right
]^{-1}=
\left(\frac{2e^4E_0^2}{3\hbar\omega m_e^2c^3}\right)^{-1}
\end{equation}
is the life time of an emitting electron in the Floquet state (otherwise, the electron-photon interaction destroys the Floquet state and cannot be considered as a weak perturbation).

Applying Eq.~\eqref{wqL}, one can find the total power that an electron radiates,
\begin{equation}\label{PL}
P_0=\sum_\mathbf{q}\hbar cqW^{(a)}_\mathbf{k}(\mathbf{q})=\frac{2}{3}\left(\frac{e^4E^2_0}{m^2_ec^3}\right),
\end{equation}
which is identical to the classical Larmor formula, $P_0=(2e^2/3c^3)\dot{\mathbf{v}}^2$, where the electron velocity $\mathbf{v}$ is defined by Eq.~\eqref{vL}.  As to the recoil force acting on the electron (the momentum transferred to the electron per unit time due to the photon emission),
\begin{equation}\label{LADL}
\mathbf{F}_{\|}=-\sum_{\mathbf{q}}\hbar \mathbf{q}W^{(a)}_\mathbf{k}(\mathbf{q})=-\frac{2}{3}\left(\frac{e^4E^2_0}{m^2_ec^5}\right)\mathbf{v}_\mathbf{k},
\end{equation}
it is identical to the classical LAD force~\cite{Barut_book},
\begin{equation}\label{ALL}
\mathbf{F}_{\mathrm{LAD}}=\left(\frac{2e^2}{3c^3}\right)\left[\gamma^2\ddot{\mathbf{v}}+
\frac{\gamma^4\mathbf{v}(\mathbf{v}\cdot
\ddot{\mathbf{v}})}{c^2}
+\frac{3\gamma^4\dot{\mathbf{v}}(\mathbf{v}\cdot
\dot{\mathbf{v}})}{c^2}\right],
\end{equation}
where $\gamma=1/\sqrt{1-{v}^2/c^2}$ is the Lorentz factor. Indeed, substituting the velocity \eqref{vL} into Eq.~\eqref{ALL} and
averaging it over the field period, $2\pi/\omega$, we arrive at the same Eq.~\eqref{LADL}. As a consequence, Eq.~\eqref{LADL} exactly coincides also with the Landau-Lifshitz equation describing the radiation reaction in an electromagnetic field~\cite{Landau_2}. To extend the developed theory for an electron in an electromagnetic wave, one needs to make the replacement $\omega t\rightarrow\omega t-\mathbf{q}_0\mathbf{r}$ in the vector potential~\eqref{A0L}, where $\mathbf{q}_0=(0,\,0,\,\omega/c)$ is the wavevector of the wave. It can be easily proved that the Floquet functions \eqref{psiL} with the same replacement stay eigenfunctions of the Hamiltonian $\hat{\cal H}_0$ in the non-relativistic limit ($v/c,\,\hbar\omega/m_ec^2\ll1$). As an only consequence of such a modification of them, the replacement of the delta function, $\delta^3(\mathbf{k}-\mathbf{k}'\pm\mathbf{q})\rightarrow\delta^3(\mathbf{k}-\mathbf{k}'\pm\mathbf{q}+m\mathbf{q}_0)$,
should be done in the matrix element~\eqref{VnmL}. Physically, this means that emission of each photon at the main radiation harmonic with $m=1$ is accompanied by transfer of the momentum $\hbar\mathbf{q}_0$ from the wave to the electron (the Compton scattering). Therefore, the photon drag force,
\begin{equation}\label{F0L}
\mathbf{F}_{0}=\hbar\mathbf{q}_0\sum_\mathbf{q}W^{(a)}_\mathbf{k}(\mathbf{q})=\frac{\hbar\mathbf{q}_0}{\tau}=\frac{2}{3}
\left(\frac{e^4E_0^2}{m^2_ec^4}\right)\mathbf{n}_0,
\end{equation}
occurs in addition to the recoil force \eqref{LADL}, where $\mathbf{n}_0=\mathbf{q}_0/q_0$ is the unit wavevector. As expected, Eq.~\eqref{F0L} is identical to the classical radiation pressure force acting on an electron under an electromagnetic wave~\cite{Landau_2,Bohm}.

It follows from the aforesaid that the one-vertex emission process pictured in Fig.~1a leads to exactly the same physical results as the classical electrodynamics. To find the first quantum correction to them, one needs to take into account the one-loop process pictured in Fig.~1b. Solving the Schr\"odinger equation \eqref{dckL} within the third order of the conventional perturbation theory~\cite{Landau_3}, we arrive at the amplitude of the one-loop process,
\begin{align}\label{ck3L}
&c^{(b)}_{\mathbf{k}'}(t')=\frac{i}{\hbar^3}\sum\limits_{\mathbf{q}'}
\sum\limits_{\mathbf{e}^{(\mathbf{q}')}}\int_0^{t'}dt
\sum\limits_{\mathbf{k}'''}
\langle\psi_{\mathbf{k}'}|\hat{V}^{(+\mathbf{q}')}|\psi_{\mathbf{k}'''}\rangle \nonumber\\
&\times\int^{t}dt
\left[\sum\limits_{\mathbf{k}''}\langle\psi_{\mathbf{k}'''}|\hat{V}^{(-\mathbf{q})}|\psi_{\mathbf{k}''}\rangle \int^{t}dt\langle\psi_{\mathbf{k}''}|\hat{V}^{(-\mathbf{q}')}|\psi_{\mathbf{k}}\rangle \right].
\end{align}
Then the total probability of photon emission per unit time reads
\begin{align}\label{wF3L}
&W_\mathbf{k}(\mathbf{q})=\sum_{\mathbf{e}^{(\mathbf{q})}}
\sum_{\mathbf{k}'}\frac{d \left|c^{(a)}_{{\mathbf{k}'}}(t')+c^{(b)}_{{\mathbf{k}'}}(t')\right|^2}{dt'}=\frac{2\pi}{\hbar}\sum_{\mathbf{e}^{(\mathbf{q})}}
\left|T_\mathbf{k}(\mathbf{q})\right|^2\nonumber\\
&\times\delta(\varepsilon_{\mathbf{k}-\mathbf{q}}-\varepsilon_{\mathbf{k}}-\hbar\omega+\hbar cq)\approx W^{(a)}_\mathbf{k}(\mathbf{q})+{w}_\mathbf{k}(\mathbf{q}),
\end{align}
where $T_\mathbf{k}(\mathbf{q})=T^{(a)}_\mathbf{k}(\mathbf{q})+T^{(b)}_\mathbf{k}(\mathbf{q})$ is the total probability amplitude of photon emission,
\begin{align}\label{T0L}
&T^{(b)}_\mathbf{k}(\mathbf{q})=
\sum\limits_{{\mathbf{q}'}}\sum\limits_{\mathbf{e}^{(\mathbf{q}')}}\sum_{m'',m'=-\infty}^{\infty}
V_{\mathbf{k}-\mathbf{q},\,\mathbf{k}-\mathbf{q}'-\mathbf{q}}^{\left(1-m''-m'\right)}V_{\mathbf{k}-\mathbf{q}'-\mathbf{q},\,\mathbf{k}-\mathbf{q}'}^{\left(m''\right)}\nonumber\\
&\times V_{\mathbf{k}-\mathbf{q}',\,\mathbf{k}}^{\left(m'\right)}\,\,
\Big(\varepsilon_\mathbf{k}-\varepsilon_{\mathbf{k}-\mathbf{q}'}+m'\hbar\omega- \hbar cq'+i0\Big)^{-1}\nonumber\\
&\times\Big(\varepsilon_\mathbf{k}-\varepsilon_{\mathbf{k}-\mathbf{q}'-\mathbf{q}}+(m'+m'')\hbar\omega- \hbar c(q'+q)+i0\Big)^{-1}
\end{align}
is the probability amplitude of the one-loop process, and
\begin{align}\label{w1L}
&{w}_\mathbf{k}(\mathbf{q})=\frac{2}{3}\left(\frac{\pi^2e^2v_0^2}
{cq\cal{V}}\right)\left(\frac{v_0}{c}\right)^2\left(\frac{e^2}{\hbar c}\right)\left(\frac{v_{\mathbf{k}y}}{c}\sin\theta_{\mathbf{q}}\cos\varphi_{\mathbf{q}}\right.\nonumber\\
&\left.-\frac{v_{\mathbf{k}x}}{c}\sin\theta_{\mathbf{q}}\sin\varphi_{\mathbf{q}}
\right)\delta(\varepsilon_{\mathbf{k}-\mathbf{q}}-\varepsilon_{\mathbf{k}}-\hbar\omega+\hbar cq)
\end{align}
is the QED correction to the probability \eqref{wqL} arisen from the one-loop process~\cite{SM}. Applying Eq.~\eqref{wF3L}, the total power emitted by an electron can be written as
\begin{align}\label{IL}
&{{P}}=\sum_{\mathbf{q}}\hbar cqW_\mathbf{k}({\mathbf{q}})=\frac{\hbar c\cal{V}}{(2\pi)^3}\int_0^{2\pi}d\varphi_{\mathbf{q}}
\int_0^{\pi}d\theta_{\mathbf{q}}\sin\theta_{\mathbf{q}}\nonumber\\
&\times\int_0^\infty dq \,q^3W_\mathbf{k}({\mathbf{q}})=
\int_0^{2\pi}d\varphi
\int_0^{\pi}d\theta\, I(\theta,\varphi)\sin\theta,
\end{align}
where
\begin{equation}\label{I1L}
I(\theta,\varphi)=\left.\frac{\hbar c\cal{V}}{(2\pi)^3}\int_0^\infty dq \,q^3W_\mathbf{k}({\mathbf{q}})\right|_{
\theta_{\mathbf{q}}=
\theta,\,
\varphi_{\mathbf{q}}=\varphi}
\end{equation}
is the angle distribution of the radiation emitted by the electron (the radiation pattern), which can be written explicitly as
\begin{align}\label{I0L}
&I(\theta,\varphi)=\frac{e^2\omega^2v^2_0}{8\pi c^3}
\Bigg[1+\cos^2+\frac{v_{\mathbf{k}x}}{c}\left(5\cos^2\theta+3\right)\sin\theta\cos\varphi\nonumber\\
&+\frac{v_{\mathbf{k}y}}{c}\left(5\cos^2\theta+3\right)\sin\theta\sin\varphi+\frac{v_{\mathbf{k}z}}{c}\left(5\cos^2\theta+1\right)\cos{\theta}
\nonumber\\
&+\frac{2}{3}
\left(\frac{v_0}{c}\right)^2\left(\frac{e^2}{\hbar c}\right)
\left(\frac{v_{\mathbf{k}y}}{c}\sin{\theta}\cos{\varphi}-
\frac{v_{\mathbf{k}x}}{c}\sin{\theta}\sin{\varphi}\right)\Bigg].
\end{align}
\begin{figure}[!h]
\includegraphics[width=0.8\linewidth]{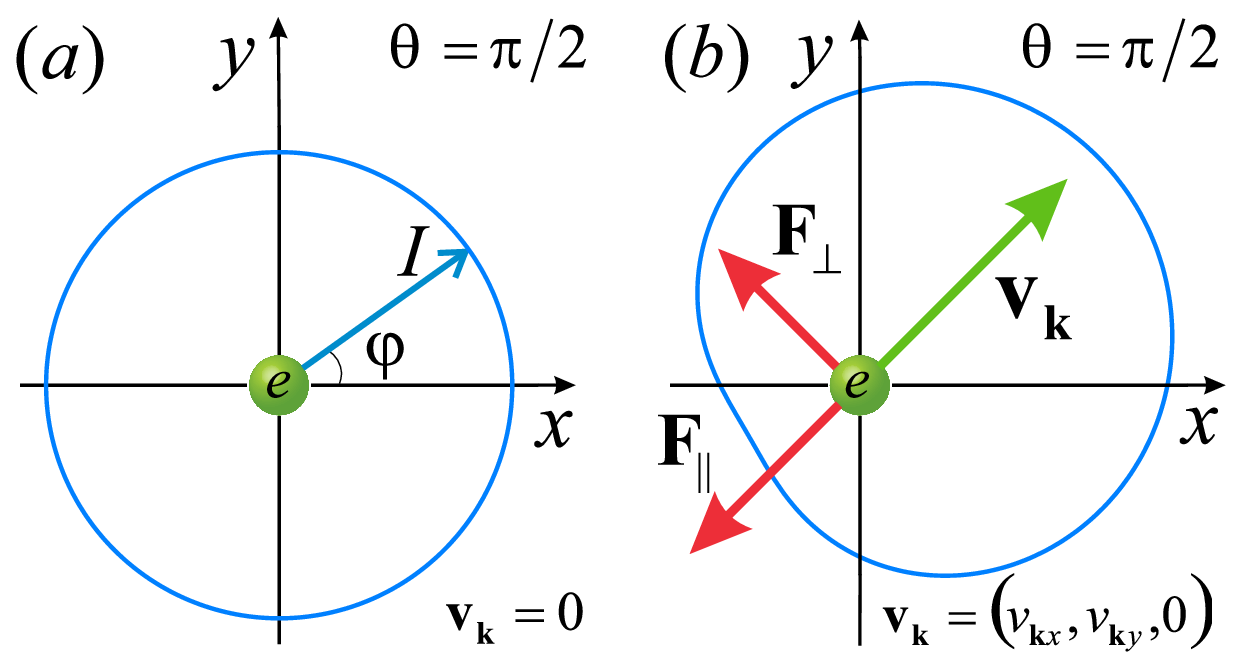}
\caption{The radiation pattern $I(\theta,\varphi)$ of an emitting electron $e$ in the $x,y$ plane: (a) for the electron at rest; (b) for the electron making forward movement with the velocity $\mathbf{v_k}$, where $\mathbf{F}_\|$ and $\mathbf{F}_\perp$ is the classical and quantum recoil force, respectively, acting on the electron due to the asymmetrical shape of the pattern.}
\end{figure}
This pattern is of symmetric shape for an electron at rest (see Fig.~2a) but it acquires asymmetry if the electron moves (see Fig.~2b). Due to such an asymmetric photon emission, the resultant momentum transferred to an emitting electron per unit time (the recoil force) differs from zero and reads
\begin{align}\label{F1L}
&\mathbf{F}=-\sum_{\mathbf{q}}\hbar\mathbf{q}W_{\mathbf{k}}({\mathbf{q}})=-
\frac{1}{c}\int_0^{2\pi}d\varphi
\int_0^{\pi}d\theta\, \mathbf{n}I(\theta,\varphi)\sin\theta\nonumber\\
&=\mathbf{F}_\|+\mathbf{F}_\perp,
\end{align}
where $\mathbf{n}=\mathbf{r}/r=(\sin\theta\cos\varphi,\,
\sin\theta\sin\varphi,\,\cos\theta)$
is the unit radius vector, $\mathbf{F}_\|$ is the normal classical recoil force \eqref{LADL} directed oppositely to the electron velocity $\mathbf{v_k}$,
\begin{equation}\label{FKL}
\mathbf{F}_\perp=-\sum_{\mathbf{q}}\hbar\mathbf{q}{w}_\mathbf{k}({\mathbf{q}})=\frac{1}{9}\left(\frac{e^4E^2_0}{m^2_ec^5}\right)
\left(\frac{v_0}{c}\right)^2\left(\frac{e^2}{\hbar c}\right)[\mathbf{L}\times\mathbf{v}_\mathbf{k}]
\end{equation}
is the anomalous recoil force directed perpendicularly to the velocity $\mathbf{v_k}$, and $\mathbf{L}$ is the unit vector along the field angular momentum, which defines the field polarization: The vector $\mathbf{L}=(0,\,0,\,1)$ describes the clockwisely polarized field \eqref{A0L}, whereas the vector $\mathbf{L}=(0,\,0,\,-1)$ corresponds to the counterclockwise field polarization.

It should be noted that the emission processes described by the loop and loopless Feynman diagrams make substantially different contributions to the radiation reaction~\eqref{F1L}. Mathematically, this follows from the fact that the anomalous recoil force~\eqref{FKL} appears due to the singularities of the probability amplitude~\eqref{T0L}, which arise from virtual photons~\cite{SM}. In contrast to the loop diagrams, the loopless ones do not involve virtual photons and, therefore, their amplitudes are devoid of the singularities. As a consequence, the two emission processes pictured in Fig.~1 describe the radiation reaction~\eqref{F1L} accurately within the leading order of electron-photon interaction: The one-vertex process makes main contribution to the normal recoil force~\eqref{LADL}, whereas the one-loop process is responsible for the anomalous force~\eqref{F1L}. It should be stressed that the one-vertex photon emission pictured in Fig.~1a is the direct emission process which does not involve intermediate electron-photon states and, therefore, is similar physically to the classical process of electromagnetic emission. As a consequence, it results in the classical recoil force~\eqref{LADL}, where the Planck constant, $\hbar$, is cancelled. As to the one-loop photon emission pictured in Fig.~1b, it is the indirect process going through the intermediate state with the virtual photon $\mathbf{q}'$. It has no analogue within classical electrodynamics and, therefore, yields the anomalous recoil force~\eqref{FKL} involving the fine structure constant, $\alpha=e^2/\hbar c$, which is of purely quantum nature.

\begin{figure}[!h]
\includegraphics[width=.8\linewidth]{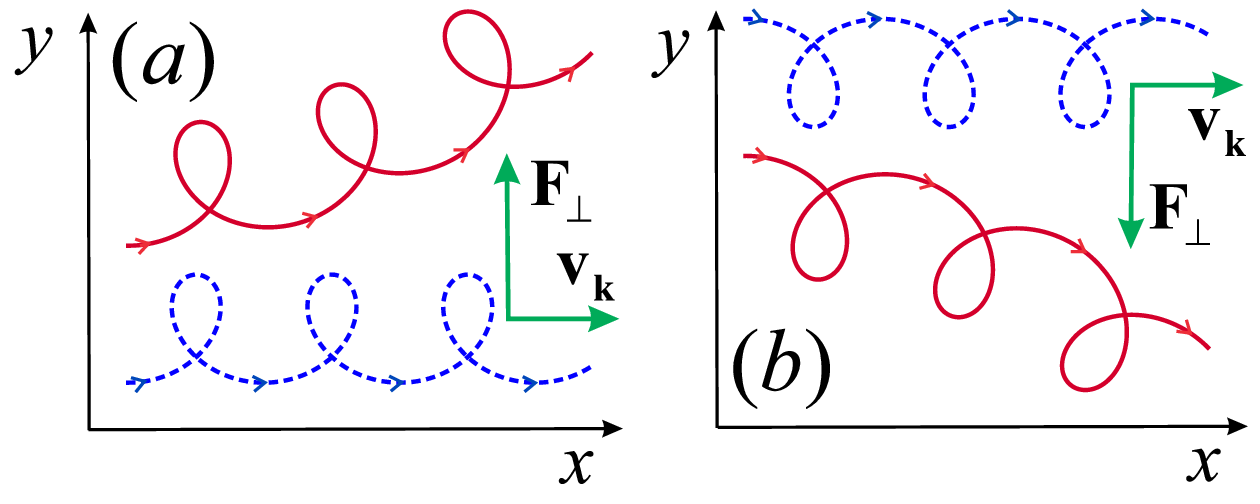}
\caption{Electron trajectories under a circularly polarized field rotating in the $x,y$ plane: (a) for clockwise field polarization; (b) for counterclockwise field polarization. The trajectories describing the electron motion within the classical electrodynamics are plotted by the dashed lines, whereas the solid lines indicate these trajectories under the quantum recoil force, $\mathbf{F}_\perp$, directed perpendicularly to the forward movement velocity, $\mathbf{v_k}$.}
\end{figure}
Evolution of the electron velocity $\mathbf{v_k}$  under the force \eqref{F1L} can be described by the Newton equation,
$
m_e\dot{\mathbf{v}}_\mathbf{k}=\mathbf{F}_\|+\mathbf{F}_\perp$,
which leads to the electron trajectory pictured schematically in Fig.~3. Since the quantum recoil force, $\mathbf{F}_\perp$, is directed perpendicularly to the electron velocity, it does not decelerate an electron in contrast to the classical recoil force, $\mathbf{F}_\|$, but bends its trajectory similarly to the Lorentz force acting on an electron in a magnetic field. Such an analogy has the deep physical nature since both a high-frequency circularly polarized field and a stationary magnetic field break the time-reversal symmetry. As a consequence, a circularly polarized field can induce various magnetic-like effects, including the spin polarization~\cite{Kibis_2022}, bandgap opening in solids~\cite{Kibis_2010,Wang_2013,Sie_2015,Cavalleri_2020}, persistent currents in conductors~\cite{Kibis_2011,Melnikov_2021}, etc. It should be stressed that the spatial asymmetry of photon emission responsible for the quantum recoil force \eqref{FKL} --- which is described by the last term in Eq.~\eqref{I0L} --- also arises from the broken time-reversal symmetry. Therefore, the discussed anomalous radiation reaction acting on a charged particle perpendicularly to its velocity takes place in any electromagnetic field with nonzero angular momentum $\mathbf{L}$ (circularly and elliptically polarized fields) and vanishes only in the case of linearly polarized field with $\mathbf{L}=0$. It follows from the aforesaid, particularly, that Eq.~\eqref{FKL} should be considered as a quantum correction to the Landau-Lifshitz equation~\cite{Landau_2} describing the classical radiation reaction force in an electromagnetic field of most general form. It should be noted also that the angular momentum of a circularly polarized field, $\mathbf{L}$,  is codirectional to the angular momentum of a charge rotating under the field. This means that the quantum recoil force \eqref{FKL} is similar phenomenologically to the classical Magnus force, $\mathbf{F}_M\propto[\mathbf{L}\times\mathbf{v}]$, acting on a body which rotates with the angular momentum $\mathbf{L}$ and makes forward movement with the velocity $\mathbf{v}$ in a medium with friction (see, e.g., Ref.~\onlinecite{Clancy_book}). Treating the classical recoil force \eqref{LADL} as a radiation friction, the quantum recoil force \eqref{FKL} can be considered qualitatively as a specific kind of the Magnus force arisen from this friction acting on a rotating charge.

Although the present theory is developed for the homogeneous field \eqref{A0L}, the force~\eqref{FKL} very weakly depends on spatial profile of the field if its characteristic scale much exceeds the radius of circular electron trajectory, $r_0=|v_0|/\omega$. As to the time profile of the field (the laser pulse duration), its characteristic scale should much exceed the radiative life time \eqref{tauL}. Since the radiative life time, $\tau$, must be large enough to satisfy the condition $\omega\tau\gg1$, it is most suitable to detect the force~\eqref{FKL} with non-relativistic laser fields which can be easily generated in the continuous-wave regime. Discussing dependence of the force~\eqref{FKL} on the field strength, $E_0$, it should be noted that the Floquet functions \eqref{psiL} are exact solutions of the non-relativistic Schr\"odinger problem for an electron in the circularly polarized background field \eqref{A0L} of any strength. Therefore, the developed theory is applicable to the background field of the strength $E_0$ satisfying only the non-relativistic condition, $v_0/c=eE_0/m_e\omega c\ll1$. To keep the ratio $v_0/c$ to be small enough within applicability of the developed non-relativistic theory, one can consider laser fields of the strength $E_0$ up to $10^{11}$~V/m. Estimating the force \eqref{FKL} quantitatively, it should be noted that the electric field amplitude $E_0$ can reach $10^{10}$~V/m for an electromagnetic wave generated by an ordinary hundred-kilowatt laser with the wavelengths $\lambda\sim10^{-6}$~m. In such a field, the quantum force \eqref{FKL} yields the electron acceleration $a_\perp=F_\perp/m_e\sim10^3(v_\mathbf{k}/c)$~m/s$^2$, which can be macroscopically large even for non-relativistic electrons satisfying the condition $v_\mathbf{k}\ll c$. Thus, the discussed anomalous radiation reaction is the rare case of macroscopic QED effect which can be observable for slow electrons in laser fields of non-relativistic intensity. As to analysis of fast electrons moving in ultra-strong fields, it needs the relativistic extension of the developed Floquet theory based on the Dirac equation~\cite{SM}.

{\it Conclusion}---It is shown theoretically that the emission of photons by an electron rotating under a high-frequency circularly polarized electromagnetic field yields the quantum recoil force acting on the electron perpendicularly to the velocity of its forward movement. In contrast to the classical Lorentz-Abraham-Dirac recoil force directed oppositely to the velocity, it does not decelerate the electron but substantially modifies its trajectory. Such an anomalous radiation reaction arises from the one-loop QED correction to the photon emission and can manifest itself in strong laser fields.

{\it Acknowledgments}---This work was supported by the Foundation
for the Advancement of Theoretical Physics and Mathematics
``BASIS'' and the Ministry of Science and Higher Education of the Russian Federation (Project FSUN-2023-0006).

\renewcommand{\theequation}{A\arabic{equation}}
\setcounter{equation}{0}
\setcounter{figure}{0}
\newpage\newpage

\widetext{

\begin{center}{\large\bf SUPPLEMENTAL MATERIAL: The article text with added details of derivations}\end{center}

\begin{center}{\Large Anomalous radiation reaction in a circularly polarized field}\end{center}

\begin{center}{\large O. V. Kibis}\end{center}

\begin{center}{\it Department of Applied and Theoretical Physics,
Novosibirsk State Technical University, Karl Marx Avenue 20,
Novosibirsk 630073, Russia}\end{center}

\section{Introduction}
The laws of nature describing the motion of charged particles in electromagnetic fields lie in the basis of most physical processes in the Universe and, therefore, they are permanently in the focus of attention despite their long prehistory.  Since a charged particle driven by a high-frequency field makes oscillating movement, it emits electromagnetic radiation. If the field is strong enough, the radiation reaction dominates the particle motion and needs the quantum electrodynamics (QED) to be described accurately. Although this research area is excited nowadays due to achievements in the laser generation of strong fields~\cite{Gonoskov_2022,Fedotov_2023,Piazza_2012,Piazza_2008,Piazza_2010_1,Piazza_2010,Piazza_2009,Ilderton_2013,Cole_2018,Poder_2018}, the features of the QED corrections depending on the field polarization still wait for detailed analysis. To fill partially this gap, the emission of photons by an electron rotating under a circularly polarized field was considered within the Floquet theory which is conventionally used to describe various periodically driven quantum systems~\cite{Goldman_2014,Bukov_2015,Eckardt_2015,Casas_2001,Kibis_2022}. Solving this problem, it was found that the photon emission leads to the quantum recoil force acting on the emitting electron perpendicularly to the velocity of its forward movement. This result is surprising since the classical Lorentz-Abraham-Dirac (LAD) force --- which was introduced into physics at the dawn of electrodynamics to describe the radiation reaction and takes deserved place in textbooks~\cite{Landau_2,Barut_book} --- is directed oppositely to the velocity. The present paper is devoted to the first theoretical analysis of this anomalous radiation reaction which can manifest itself in strong laser fields.

\section{Model}
For definiteness, let us consider a classically strong circularly polarized homogeneous electromagnetic field with the zero scalar potential and the vector potential
\begin{equation}\label{A0}
\mathbf{A}=(A_x,A_y,A_z)=\left(\frac{cE_0}{\omega}\right)(\cos\omega t,\,\sin\omega t,\,0),
\end{equation}
where $E_0$ is the electric field amplitude, $\omega$ is the field frequency, $c$ is the light speed in vacuum, and the field polarization is assumed to be clockwise. Within the framework of classical electrodynamics, the vector potential \eqref{A0} describes the rotating electric field, $\mathbf{E}=-(1/c)\partial_t\mathbf{A}$, which induces electron rotation with the velocity $v_0=eE_0/m_e\omega$ along circular trajectory of the radius $r_0=|v_0|/\omega$, where $m_e$ is the electron mass and $e$ is the electron charge~\cite{Landau_2}. Proper generalization of the model field \eqref{A0} to the physically important case of electromagnetic wave will be done in the following. The non-relativistic quantum dynamics of an electron under the field \eqref{A0} is defined by the Hamiltonian
\begin{equation}\label{H0}
\hat{\cal H}_0=\frac{1}{2m_e}\left(\hat{\mathbf{p}}-\frac{e}{c}\mathbf{A}\right)^2,
\end{equation}
where $\hat{\mathbf{p}}=(\hat{{p}}_x,\,\hat{{p}}_y,\,\hat{{p}}_z)$
is the momentum operator. The exact eigenfunctions of the Hamiltonian \eqref{H0} are
\begin{equation}\label{psi}
\psi_\mathbf{k}=\frac{1}{\sqrt{\cal{V}}}e^{i\mathbf{k}\mathbf{r}}
e^{-i(\varepsilon_\mathbf{k}
+\varepsilon_0)t/\hbar}e^{i(v_0/\omega)k_x\sin\omega t}e^{-i(v_0/\omega)k_y\cos\omega t}=\frac{1}{\sqrt{\cal{V}}}e^{i\mathbf{k}\mathbf{r}}e^{-i(\varepsilon_\mathbf{k}
+\varepsilon_0)t/\hbar}e^{-i(kv_0/\omega)\sin\theta_\mathbf{k}\sin(\varphi_\mathbf{k}-\omega t)},
\end{equation}
where $\mathbf{r}=(x,y,z)=(r\sin\theta\cos\varphi,\,r\sin\theta\sin\varphi,\,r\cos\theta)$ is the electron radius vector, $\theta$ and $\varphi$ are the polar and azimuth angle, respectively, $\mathbf{k}=(k_x,\,k_y,\,k_z)=
(k\sin\theta_\mathbf{k}\cos\varphi_\mathbf{k},\,k\sin\theta_\mathbf{k}\sin\varphi_\mathbf{k},\,k\cos\theta_\mathbf{k})$ is the electron wave vector, $\varepsilon_\mathbf{k}=\hbar^2{\mathbf{k}}^2/2m_e$ is the kinetic energy of forward movement of the electron, $\varepsilon_0=m_ev_0^2/2$ is the kinetic energy of electron rotation under the field \eqref{A0}, and $\cal{V}$ is the normalization volume.
The wave functions \eqref{psi} can be easily proved by direct substitution into the Schr\"odinger equation, $\hat{\cal H}_0\psi_{\mathbf{k}}=i\hbar\partial_t\psi_{\mathbf{k}}$, and should be treated as the Floquet functions~\cite{Goldman_2014,Bukov_2015,Eckardt_2015,Casas_2001} describing dynamics of an electron strongly coupled to the field \eqref{A0} (dressed by the field). The averaged electron velocity in the Floquet state \eqref{psi} reads
\begin{equation}\label{v}
{\mathbf{v}}=\frac{i}{\hbar}\langle\psi_\mathbf{k}|\hat{\cal H}_0\mathbf{r}-\mathbf{r}\hat{\cal H}_0|\psi_\mathbf{k}\rangle=\mathbf{v_k}+\mathbf{v}_0,
\end{equation}
where $\mathbf{v_k}=\hbar\mathbf{k}/m_e$ and $\mathbf{v}_0=v_0(-\cos\omega t,\,-\sin\omega t,\,0)$ are the velocity of forward and rotational movement of the electron, respectively. As expected, the velocity \eqref{v} exactly coincides with the classical velocity of an electron under a circularly polarized field~\cite{Landau_2}. Next, let us find corrections to this velocity arisen from the emission of photons by the dressed electron.

The interaction between an electron and a photon in the presence of the background field \eqref{A0} is described by the Hamiltonian $\hat{\cal H}=\hat{\cal H}_0+\hat{V}^{(\pm\mathbf{q})}$, where $\hat{V}^{(\pm\mathbf{q})}$ is the Hamiltonian of electron-photon interaction, and the signs ``$+$'' and ``$-$'' correspond to the absorption and emission of a photon with the wave vector $\mathbf{q}$, respectively. Since the Floquet functions \eqref{psi} form the complete orthonormal function system for any time $t$, the solution of the Schr\"odinger problem with the Hamiltonian $\hat{\cal H}$ can be sought as an expansion
\begin{equation}\label{exp}
\psi=\sum_{\mathbf{k}'}c_{\mathbf{k}'}(t)\psi_{\mathbf{k}'}.
\end{equation}
Substituting the expansion \eqref{exp} into the Schr\"odinger equation, $\hat{\cal H}\psi=i\hbar\partial_t\psi$, we arrive at the Schr\"odinger equation written in the Floquet representation,
\begin{equation}\label{dck}
i\hbar\frac{dc_{\mathbf{k}'}(t)}{dt}=\sum_{\mathbf{k}}\langle\psi_{\mathbf{k}'}|\hat{V}^{(\pm\mathbf{q})}|\psi_{\mathbf{k}}\rangle c_{\mathbf{k}}(t),
\end{equation}
which defines the expansion coefficients $c_{\mathbf{k}}(t)$. Within the conventional quantum electrodynamics~\cite{Landau_4}, the matrix element of the electron interaction with a single photon reads
\begin{equation}\label{V0}
\langle\psi_{\mathbf{k}'}|\hat{V}^{(\pm\mathbf{q})}|\psi_{\mathbf{k}}\rangle=\frac{e}{c}
\int_{\cal{V}}\left(\mathbf{j}_{\mathbf{k}'\mathbf{k}}\cdot\mathbf{A}^{(\pm\mathbf{q})}\right)d^3\mathbf{r},
\end{equation}
where $\mathbf{A}^{(\pm\mathbf{q})}=\mathbf{e}^{(\mathbf{\pm q})}\sqrt{2\pi\hbar c/q\cal{V}}\,e^{\pm i(\mathbf{qr}-cqt)}$ is the photon wave function, $\mathbf{q}=(q_x,\,q_y,\,q_z)=(q\sin\theta_\mathbf{q}\cos\varphi_\mathbf{q},\,q\sin\theta_\mathbf{q}\sin\varphi_\mathbf{q},\,q\cos\theta_\mathbf{q})$ is the photon wave vector, ${\mathbf{e}^{(\mathbf{+q})}}^\ast={\mathbf{e}^{(\mathbf{-q})}}\equiv\mathbf{e}^{(\mathbf{q})}$ is the unit vector of photon polarization,
\begin{equation}\label{jk}
\mathbf{j}_{\mathbf{k}'\mathbf{k}}=\frac{1}{2m_e}(\psi^\ast_{\mathbf{k}'}\hat{\mathbf{p}}\psi_{\mathbf{k}}-
\psi_{\mathbf{k}}\hat{\mathbf{p}}\psi^\ast_{\mathbf{k}'})-\frac{e}{m_ec}\mathbf{A}
\psi^\ast_{\mathbf{k}'}\psi_{\mathbf{k}}+{\it O}(1/c^2)=\frac{e^{i(\mathbf{k}-\mathbf{k}')\mathbf{r}}}{\cal{V}}
\sum_{m=-\infty}^{\infty}\mathbf{j}_{\mathbf{k}'\mathbf{k}}^{(m)}e^{-im\omega t}
\end{equation}
is the transition current, and $\mathbf{j}_{\mathbf{k}'\mathbf{k}}^{(m)}$ are its Fourier harmonics. Applying the Jacobi-Anger expansion, $e^{i\xi\sin\omega t}=\sum_{m=-\infty}^{\infty}J_m(\xi)e^{im\omega t}$, to find these harmonics, the matrix element \eqref{V0} can be rewritten as
\begin{equation}\label{Vnm}
\langle\psi_{\mathbf{k}'}|\hat{V}^{(\pm\mathbf{q})}|\psi_{\mathbf{k}}\rangle=
\frac{(2\pi)^3}{\cal{V}}
e^{-i(\varepsilon_\mathbf{k}-\varepsilon_{\mathbf{k}\pm\mathbf{q}}\pm \hbar cq)t/\hbar}\sum_{m=-\infty}^{\infty}
V_{\mathbf{k}\pm\mathbf{q},\,\mathbf{k}}^{(m)}
e^{-im\omega t}\delta^3(\mathbf{k}-\mathbf{k}'\pm\mathbf{q}),
\end{equation}
where
\begin{align}\label{Vmm}
&V_{\mathbf{k}\pm\mathbf{q},\,\mathbf{k}}^{(m)}=e\sqrt{\frac{2\pi\hbar }{cq\cal{V}}}
\left[\left(\mathbf{v_k}\cdot\mathbf{e}^{(\mathbf{\pm q})}\right)J_{m}(\xi_\mathbf{q}^\pm)e^{im\varphi_\mathbf{q}}-
v_0\frac{{e}^{(\pm \mathbf{q})}_x- i{e}^{(\pm \mathbf{q})}_y}{2}
J_{m+1}(\xi_\mathbf{q}^\pm)e^{i(m+1)\varphi_\mathbf{q}}\right.\nonumber\\
&\left.-
v_0\frac{{e}^{(\pm \mathbf{q})}_x+i{e}^{(\pm \mathbf{q})}_y}{2}
J_{m-1}(\xi_\mathbf{q}^\pm)e^{i(m-1)\varphi_\mathbf{q}}+{\it O}(1/c^2)\right].
\end{align}
$J_m(\xi_\mathbf{q}^{\pm})$ is the Bessel function of the first kind, and its argument is $\xi_\mathbf{q}^\pm=\pm(v_0/\omega)q\sin\theta_\mathbf{q}$. To keep the matrix element \eqref{Vmm} within applicability limits of the present non-relativistic theory, the Bessel functions there will be expanded in powers of $1/c$ up to terms $\sim1/c$ in the following.

\section{Results and Discussion}
\begin{figure}[!h]
\includegraphics[width=.5\linewidth]{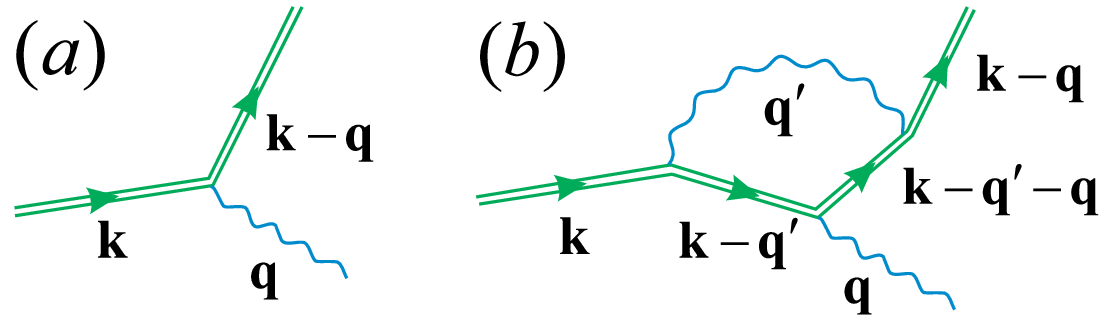}
\caption{Diagram representation of the emission of a photon with the wave vector $\mathbf{q}$ by an electron with the wave vector $\mathbf{k}$ in a background circularly polarized field: (a) the one-vertex emission process; (b) the one-loop emission process through the intermediate state of a virtual photon with the wave vector $\mathbf{q}'$. The double fermion lines indicate the electron strongly coupled to the background field.}
\end{figure}
Using the Floquet functions \eqref{psi} as a basis of the expansion \eqref{exp}, we take into account the electron interaction with the strong background field \eqref{A0} accurately (non-perturbatively), whereas the electron interaction with photons can be considered as a weak perturbation. First of all, let us consider the one-vertex photon emission pictured in Fig.~1a, assuming an electron to be in the Floquet state $\psi_{\mathbf{k}}$ at the initial time $t=0$. Solving the Schr\"odinger equation \eqref{dck} within the first order of the conventional perturbation theory~\cite{Landau_3}, we arrive at the expansion coefficient $c_{\mathbf{k}'}(t)=c^{(a)}_{\mathbf{k}'}(t)$, which describes the amplitude of the one-vertex process,
\begin{equation}\label{ck}
c^{(a)}_{\mathbf{k}'}(t')=-\frac{i}{\hbar}\int_0^{t'}\langle\psi_{\mathbf{k}'}|\hat{V}^{(-\mathbf{q})}|\psi_{\mathbf{k}}\rangle dt=-i\frac{(2\pi)^3}{\hbar\cal{V}}\sum_{m=-\infty}^{\infty}
V_{\mathbf{k}-\mathbf{q},\,\mathbf{k}}^{(m)}
\delta^3(\mathbf{k}-\mathbf{k}'-\mathbf{q})\int_0^{t'}
e^{-i(\varepsilon_\mathbf{k}-\varepsilon_{\mathbf{k}-\mathbf{q}}+m\hbar\omega- \hbar cq)t/\hbar}dt.
\end{equation}
Correspondingly, the squared modulus of Eq.~\eqref{ck},
\begin{equation}\label{sqr}
\left|c^{(a)}_{\mathbf{k}'}(t')\right|^2=\frac{(2\pi)^6}{\hbar^2{\cal{V}}^2}\left|
\sum_{m=-\infty}^{\infty}
V_{\mathbf{k}-\mathbf{q},\,\mathbf{k}}^{(m)}\delta^3(\mathbf{k}-\mathbf{k}'-\mathbf{q})
e^{-im\omega t'/2}
\int_{-t'/2}^{t'/2}e^{i(\varepsilon_{\mathbf{k}-\mathbf{q}}-\varepsilon_{\mathbf{k}}-m\hbar\omega+\hbar cq)t/\hbar}dt\right|^2,
\end{equation}
is the probability of the one-vertex photon emission during the time $t'$. In the limiting case of $t'\rightarrow\infty$, it reads
\begin{equation}\label{Sqr}
\left|c^{(a)}_{\mathbf{k}'}(t')\right|^2=\frac{(2\pi)^8}{{\cal{V}}^2}\sum_{m=-\infty}^{\infty}
\left|V_{\mathbf{k}-\mathbf{q},\,\mathbf{k}}^{(m)}\right|^2
\left[\delta^3(\mathbf{k}-\mathbf{k}'-\mathbf{q})\right]^2
\left[\delta(\varepsilon_{\mathbf{k}-\mathbf{q}}-\varepsilon_{\mathbf{k}}-m\hbar\omega+\hbar cq)\right]^2.
\end{equation}
To rewrite squared delta functions in Eq.~\eqref{Sqr} appropriately, one has to apply the transformation~\cite{Landau_4}
$$[\delta(\varepsilon)]^2=\delta(\varepsilon)\delta(0)
=\frac{\delta(\varepsilon)}{2\pi\hbar}\lim_{t'\rightarrow\infty}
\int_{-t'/2}^{t'/2}e^{i0\times
t/\hbar}dt=\frac{\delta(\varepsilon)t'}{2\pi\hbar},\,\,\,
[\delta^3(\mathbf{k})]^2=\delta^3(\mathbf{k})\delta^3(0)
=\frac{\delta^3(\mathbf{k})}{(2\pi)^3}
\int_{\cal{V}}e^{i0\times
\mathbf{r}}d^3\mathbf{r}=\frac{\delta^3(\mathbf{k})\cal{V}}{(2\pi)^3}.$$
Then Eq.~\eqref{Sqr} yields the probability of photon emission per unit time,
\begin{equation}\label{wq}
W^{(a)}_\mathbf{k}(\mathbf{q})=\sum_{\mathbf{e}^{(\mathbf{q})}}\sum_{\mathbf{k}'\neq\mathbf{k}}\frac{d \left|c^{(a)}_{\mathbf{k}'}(t')\right|^2}{dt'}=\frac{2\pi}{\hbar}\sum_{\mathbf{e}^{(\mathbf{q})}}
\sum_{m=1}^{\infty}
\left|V_{\mathbf{k}-\mathbf{q},\,\mathbf{k}}^{(m)}\right|^2
\delta(\varepsilon_{\mathbf{k}-\mathbf{q}}-\varepsilon_{\mathbf{k}}-m\hbar\omega+\hbar cq).
\end{equation}
The delta functions in Eq.~\eqref{wq} describe the momentum-energy conservation law,
$\varepsilon_{\mathbf{k}}+m\hbar\omega=\varepsilon_{\mathbf{k}-\mathbf{q}}+\hbar cq$, which yields the allowed lengths of wave vectors of photons,
\begin{equation}\label{q}
q_m=\frac{1}{\lambda_0}\left(\sqrt{[1-\left(\mathbf{v_k}\cdot
\mathbf{n}^{(\mathbf{q})}
\right)/c]^2+2m\lambda_0\omega/c}-
[1-(\mathbf{v_k}\cdot\mathbf{n}^{(\mathbf{q})})/c]\right)=
\frac{m\omega}{c}\left[1+\frac{\left(\mathbf{v_k}\cdot
\mathbf{n}^{(\mathbf{q})}\right)}{c}+{\it O}(1/c^2)\right],
\end{equation}
where $\mathbf{n}^{(\mathbf{q})}=\mathbf{q}/q=(\sin\theta_{\mathbf{q}}\cos\varphi_{\mathbf{q}},\,
\sin\theta_{\mathbf{q}}\sin\varphi_{\mathbf{q}},\,\cos\theta_{\mathbf{q}})$
is the unit wave vector, $\lambda_0=\hbar/m_ec$ is the Compton wavelength, and $m=1,2,...$ is the number of radiation harmonic.  Expanding Eq.~\eqref{Vmm} in powers of $1/c$ and taking into account Eq.~\eqref{q}, one can see that main contribution to the probability~\eqref{wq} arises from the first harmonic, $\left|V_{\mathbf{k}-\mathbf{q},\,\mathbf{k}}^{(1)}\right|^2$,
whereas contribution of higher harmonics is relativistically small, $\left|V_{\mathbf{k}-\mathbf{q},\,\mathbf{k}}^{(m\neq1)}\right|^2\sim {\it{O}}\left({1}/{c^2}\right)$, and can be neglected as a first approximation. This agrees with the classical description of the considered emission problem: Indeed, it is known that a rotating non-relativistic charge emits electromagnetic waves (the cyclotron radiation) mainly at the rotation frequency, $\omega$, whereas the emission at multiple frequencies is relativistically weak~\cite{Landau_2}.
Thus, the probability \eqref{wq} takes the form of the Fermi's golden rule,
\begin{equation}\label{wF}
W^{(a)}_\mathbf{k}(\mathbf{q})=\frac{2\pi}{\hbar}\sum_{\mathbf{e}^{(\mathbf{q})}}
\left|T^{(a)}_\mathbf{k}(\mathbf{q})\right|^2
\delta(\varepsilon_{\mathbf{k}-\mathbf{q}}-\varepsilon_{\mathbf{k}}-\hbar\omega+\hbar cq),
\end{equation}
where
\begin{equation}\label{Ta}
T^{(a)}_\mathbf{k}(\mathbf{q})=V_{\mathbf{k}-\mathbf{q},\,\mathbf{k}}^{(1)}=
-{ev_0}\sqrt{\frac{\pi\hbar }{2cq\cal{V}}}\left[
{{e}^{(\mathbf{q})}_x+i{e}^{(\mathbf{q})}_y}+\frac{\left(\mathbf{v_k}\cdot
\mathbf{e}^{(\mathbf{q})}\right)}{c}\left({n}^{(\mathbf{q})}_x+i{n}^{(\mathbf{q})}_y\right)+{\it O}(1/c^2)\right]
\end{equation}
is the probability amplitude of the one-vertex emission process pictured in Fig.~1a. As to the summation over photon polarizations, it is effected conventionally by averaging over the directions of $\mathbf{e}^{(\mathbf{q})}$ (in a plane perpendicular to the given direction $\mathbf{n}^{(\mathbf{q})}$) and the result is then doubled (because of the two independent possible polarizations of photon), which leads to the known equality, $\sum_{\mathbf{e}^{(\mathbf{q})}}{e}^{(\mathbf{q})}_i{{e}^{(\mathbf{q})}_j}^\ast=\delta_{ij}-
{n}^{(\mathbf{q})}_i{n}^{(\mathbf{q})}_j$ (see, e.g., Sec.~45 in Ref.~\onlinecite{Landau_4}). For the coordinates used in Eq.~\eqref{Vmm}, this equality can be written as
\begin{eqnarray}\label{gr}
&\sum\limits_{\mathbf{e}^{(\mathbf{q})}}{e}^{(\mathbf{q})}_x{{e}^{(\mathbf{q})}_x}^\ast=(1-\sin^2\theta_{\mathbf{q}}
\cos^2\varphi_{\mathbf{q}}),\,\,
\sum\limits_{\mathbf{e}^{(\mathbf{q})}}{e}^{(\mathbf{q})}_y{{e}^{(\mathbf{q})}_y}^\ast=(1-\sin^2\theta_{\mathbf{q}}
\sin^2\varphi_{\mathbf{q}}),\,\,
\sum\limits_{\mathbf{e}^{(\mathbf{q})}}{e}^{(\mathbf{q})}_z{{e}^{(\mathbf{q})}_z}^\ast=\sin^2\theta_{\mathbf{q}}
,\,\,\nonumber\\
&\sum\limits_{\mathbf{e}^{(\mathbf{q})}}{e}^{(\mathbf{q})}_x{{e}^{(\mathbf{q})}_y}^\ast=
-(\sin^2\theta_{\mathbf{q}}\sin2\varphi_{\mathbf{q}})/2,\,\,
\sum\limits_{\mathbf{e}^{(\mathbf{q})}}{e}^{(\mathbf{q})}_x{{e}^{(\mathbf{q})}_z}^\ast=
-(\sin2\theta_{\mathbf{q}}\cos\varphi_{\mathbf{q}})/2,\,\,
\sum\limits_{\mathbf{e}^{(\mathbf{q})}}{e}^{(\mathbf{q})}_y{{e}^{(\mathbf{q})}_z}^\ast=
-(\sin2\theta_{\mathbf{q}}\sin\varphi_{\mathbf{q}})/2,\nonumber\\
\end{eqnarray}
which allows to write the probability~\eqref{wF} explicitly.
As a result, Eq.~\eqref{wF} takes its final form
\begin{align}\label{w}
&W^{(a)}_\mathbf{k}(\mathbf{q})=
\left[1+
\cos^2\theta_{\mathbf{q}}+\frac{2v_{\mathbf{k}x}}{c}\cos^2\theta_{\mathbf{q}}\sin\theta_{\mathbf{q}}\cos\varphi_{\mathbf{q}}+
\frac{2v_{\mathbf{k}y}}{c}\cos^2\theta_{\mathbf{q}}\sin\theta_{\mathbf{q}}\sin\varphi_{\mathbf{q}}
-\frac{2v_{\mathbf{k}z}}{c}\cos\theta_{\mathbf{q}}\sin^2{\theta_{\mathbf{q}}}+{\it O}(1/c^2)\right]\nonumber\\
&\times\frac{\pi^2e^2v_0^2}{\hbar c\omega\cal{V}}\delta(q-q_1).
\end{align}
In the following, we will assume that the field frequency, $\omega$, is high enough to satisfy the condition $\omega\tau\gg1$,
where
\begin{equation}\label{tau}
{\tau}=\left[\sum_\mathbf{q}W^{(a)}_\mathbf{k}(\mathbf{q})\right
]^{-1}=
\left[
\frac{\cal{V}}{(2\pi)^3}\int_0^{2\pi}d\varphi_{\mathbf{q}}
\int_0^{\pi}d\theta_{\mathbf{q}}\sin\theta_{\mathbf{q}}\int_0^\infty dq\,q^2\,W^{(a)}_\mathbf{k}(\mathbf{q})\right]^{-1}=
\left(\frac{2e^4E_0^2}{3\hbar\omega m_e^2c^3}\right)^{-1}
\end{equation}
is the life time of an emitting electron in the Floquet state (otherwise, the electron-photon interaction destroys the Floquet state and cannot be considered as a weak perturbation).

Applying Eq.~\eqref{w}, the total power that an electron radiates can be written as
\begin{equation}\label{P}
P_0=\sum_\mathbf{q}\hbar cqW^{(a)}_\mathbf{k}(\mathbf{q})=\frac{\hbar c\cal{V}}{(2\pi)^3}\int_0^{2\pi}d\varphi_{\mathbf{q}}
\int_0^{\pi}d\theta_{\mathbf{q}}\sin\theta_{\mathbf{q}}\int_0^\infty dq \,q^3W^{(a)}_\mathbf{k}(\mathbf{q})=\frac{2}{3}\left(\frac{e^4E^2_0}{m^2_ec^3}\right),
\end{equation}
which is identical to the classical Larmor formula, $P_0=(2e^2/3c^3)\dot{\mathbf{v}}^2$, where the electron velocity $\mathbf{v}$ is defined by Eq.~\eqref{v}. As to the recoil force acting on the electron (the momentum transferred to the electron per unit time due to the photon emission),
\begin{equation}\label{LAD}
\mathbf{F}_{\|}=-\sum_{\mathbf{q}}\hbar \mathbf{q}W^{(a)}_\mathbf{k}(\mathbf{q})=-\frac{\hbar c\cal{V}}{(2\pi)^3}\int_0^{2\pi}d\varphi_{\mathbf{q}}
\int_0^{\pi}d\theta_{\mathbf{q}}\,\mathbf{n}^{(\mathbf{q})}\sin\theta_{\mathbf{q}}
\int_0^\infty dq \,q^3W^{(a)}_\mathbf{k}(\mathbf{q})=-\frac{2}{3}\left(\frac{e^4E^2_0}{m^2_ec^5}\right)\mathbf{v}_\mathbf{k},
\end{equation}
it is identical to the classical LAD force~\cite{Barut_book},
\begin{equation}\label{AL}
\mathbf{F}_{\mathrm{LAD}}=\left(\frac{2e^2}{3c^3}\right)\left[\gamma^2\ddot{\mathbf{v}}+
\frac{\gamma^4\mathbf{v}(\mathbf{v}\cdot
\ddot{\mathbf{v}})}{c^2}
+\frac{3\gamma^4\dot{\mathbf{v}}(\mathbf{v}\cdot
\dot{\mathbf{v}})}{c^2}\right],
\end{equation}
where $\gamma=1/\sqrt{1-{v}^2/c^2}$ is the Lorentz factor. Indeed, substituting the velocity \eqref{v} into Eq.~\eqref{AL} and
averaging it over the field period, $2\pi/\omega$, we arrive at the same Eq.~\eqref{LAD}. As a consequence, Eq.~\eqref{LAD} exactly coincides also with the Landau-Lifshitz equation describing the classical radiation reaction in an electromagnetic field~\cite{Landau_2}. To extend the developed theory for an electron in an electromagnetic wave, one needs to make the replacement $\omega t\rightarrow\omega t-\mathbf{q}_0\mathbf{r}$ in the vector potential~\eqref{A0}, where $\mathbf{q}_0=(0,\,0,\,\omega/c)$ is the wavevector of the wave. It can be easily proved that the Floquet functions \eqref{psi} with the same replacement stay eigenfunctions of the Hamiltonian \eqref{H0} in the non-relativistic limit ($v/c,\,\hbar\omega/m_ec^2\ll1$). As an only consequence of such a modification of them, the replacement of the delta function, $\delta^3(\mathbf{k}-\mathbf{k}'\pm\mathbf{q})\rightarrow\delta^3(\mathbf{k}-\mathbf{k}'\pm\mathbf{q}+m\mathbf{q}_0)$,
should be done in the matrix element~\eqref{Vnm}. Physically, this means that emission of each photon at the main radiation harmonic with $m=1$ is accompanied by transfer of the momentum $\hbar\mathbf{q}_0$ from the wave to the electron (the Compton scattering). Therefore, the photon drag force,
\begin{equation}\label{F0}
\mathbf{F}_{0}=\hbar\mathbf{q}_0\sum_\mathbf{q}W^{(a)}_\mathbf{k}(\mathbf{q})=\frac{\hbar\mathbf{q}_0}{\tau}=\frac{2}{3}
\left(\frac{e^4E_0^2}{m^2_ec^4}\right)\mathbf{n}_0,
\end{equation}
occurs in addition to the recoil force \eqref{LAD}, where $\mathbf{n}_0=\mathbf{q}_0/q_0$ is the unit wavevector. As expected, Eq.~\eqref{F0} is identical to the classical radiation pressure force acting on an electron under an electromagnetic wave~\cite{Landau_2,Bohm}.

It follows from the aforesaid that the one-vertex emission process pictured in Fig.~1a leads to exactly the same physical results as the classical electrodynamics. To find the first quantum correction to them, one needs to take into account the one-loop process pictured in Fig.~1b. Solving the Schr\"odinger equation \eqref{dck} within the third order of the conventional perturbation theory~\cite{Landau_3}, we arrive at the expansion coefficient, $c_{\mathbf{k}'}(t)=c^{(b)}_{\mathbf{k}'}(t)$, describing the amplitude of this process,
\begin{align}\label{ck3}
&c^{(b)}_{\mathbf{k}'}(t')=\frac{i}{\hbar^3}\sum\limits_{\mathbf{q}'}
\sum\limits_{\mathbf{e}^{(\mathbf{q}')}}\int_0^{t'}dt
\sum\limits_{\mathbf{k}'''}
\langle\psi_{\mathbf{k}'}|\hat{V}^{(+\mathbf{q}')}|\psi_{\mathbf{k}'''}\rangle \int^{t}dt
\left[\sum\limits_{\mathbf{k}''}\langle\psi_{\mathbf{k}'''}|\hat{V}^{(-\mathbf{q})}|\psi_{\mathbf{k}''}\rangle \int^{t}dt\langle\psi_{\mathbf{k}''}|\hat{V}^{(-\mathbf{q}')}|\psi_{\mathbf{k}}\rangle \right]\nonumber\\
&=-i
\frac{(2\pi)^3}{\hbar\cal{V}}\sum_{m=-\infty}^{\infty}\delta^3(\mathbf{k}-\mathbf{k}'-\mathbf{q})
\int_0^{t'}
e^{-i(\varepsilon_\mathbf{k}-\varepsilon_{\mathbf{k}-\mathbf{q}}+m\hbar\omega- \hbar cq)t/\hbar}dt\nonumber\\
&\times
\sum\limits_{\mathbf{q}'}\sum\limits_{\mathbf{e}^{(\mathbf{q}')}}\sum_{m'',m'=-\infty}^{\infty}
\frac{V_{\mathbf{k}-\mathbf{q},\,\mathbf{k}-\mathbf{q}'-\mathbf{q}}^{\left(m-m''-m'\right)}V_{\mathbf{k}-\mathbf{q}'-\mathbf{q},\,\mathbf{k}-\mathbf{q}'}^{\left(m''\right)}V_{\mathbf{k}-\mathbf{q}',\,\mathbf{k}}^{\left(m'\right)}}
{[\varepsilon_\mathbf{k}-\varepsilon_{\mathbf{k}-\mathbf{q}'-\mathbf{q}}+(m'+m'')\hbar\omega- \hbar c(q'+q)+i0][\varepsilon_\mathbf{k}-\varepsilon_{\mathbf{k}-\mathbf{q}'}+m'\hbar\omega- \hbar cq'+i0]}.
\end{align}
Then the total probability of photon emission per unit time reads
\begin{equation}\label{wF3}
W_\mathbf{k}(\mathbf{q})=\sum_{\mathbf{e}^{(\mathbf{q})}}\sum_{\mathbf{k}'\neq\mathbf{k}}\frac{d \left|c^{(a)}_{\mathbf{k}'}(t')+c^{(b)}_{\mathbf{k}'}(t')\right|^2}{dt'}=\frac{2\pi}{\hbar}\sum_{\mathbf{e}^{(\mathbf{q})}}
\left|T^{(a)}_\mathbf{k}(\mathbf{q})+T^{(b)}_\mathbf{k}(\mathbf{q})\right|^2
\delta(\varepsilon_{\mathbf{k}-\mathbf{q}}-\varepsilon_{\mathbf{k}}-\hbar\omega+\hbar cq),
\end{equation}
where
\begin{equation}\label{T0}
T^{(b)}_\mathbf{k}(\mathbf{q})=
\int_{-\infty}^\infty \frac{{\cal{V}}d^3\mathbf{q}'}{(2\pi)^3}\sum\limits_{\mathbf{e}^{(\mathbf{q}')}}\sum_{m'',m'=-\infty}^{\infty}
\frac{V_{\mathbf{k}-\mathbf{q},\,\mathbf{k}-\mathbf{q}'-\mathbf{q}}^{\left(1-m''-m'\right)}V_{\mathbf{k}-\mathbf{q}'-\mathbf{q},\,\mathbf{k}-\mathbf{q}'}^{\left(m''\right)}V_{\mathbf{k}-\mathbf{q}',\,\mathbf{k}}^{\left(m'\right)}}
{[\varepsilon_\mathbf{k}-\varepsilon_{\mathbf{k}-\mathbf{q}'-\mathbf{q}}+(m'+m'')\hbar\omega- \hbar c(q'+q)+i0][\varepsilon_\mathbf{k}-\varepsilon_{\mathbf{k}-\mathbf{q}'}+m'\hbar\omega- \hbar cq'+i0]}
\end{equation}
is the probability amplitude of the one-loop emission process. It should be noted that the integrand of Eq.~\eqref{T0} has singularities arisen from virtual photons with the wave vectors $\mathbf{q}'$ satisfying the momentum-energy conservation law (the on-shell photons). To carry out the integration over all photon wave vectors $\mathbf{q}'$ accurately, one needs to apply the conventionally used Sokhotski–Plemelj formula,
\begin{equation}\label{T1}
\frac{1}{x-a-i0}={\cal{P}}\frac{1}{x-a}+i\pi\delta(x-a),
\end{equation}
where the symbol ${\cal{P}}$ denotes the principal value of the integral~(see, e.g., Sec.~43 in Ref.~\onlinecite{Landau_3}). Calculating the first term in the right-hand side of Eq.~\eqref{T1}, one can see that it makes only the relativistically small contribution to the classical recoil force \eqref{LAD} and, therefore, will be neglected in the following analysis. As to the second delta-function term accounting for the on-shell photons, it yields the substantially new addition ${w}_\mathbf{k}(\mathbf{q})\propto v_{{\mathbf{k}}y,x}q_{x,y}$ to the probability \eqref{w}, which crucially modifies the electron dynamics discussed below. Keeping in mind that an electron is assumed to be slow enough ($v_\mathbf{k}\ll v_0\ll c$), the total probability amplitude, $T_\mathbf{k}(\mathbf{q})=T^{(a)}_\mathbf{k}(\mathbf{q})+T^{(b)}_\mathbf{k}(\mathbf{q})$, can be written in the leading order of expansion in powers of $1/c$ as
\begin{align}\label{T00}
&T_\mathbf{k}(\mathbf{q})\approx V_{\mathbf{k}-\mathbf{q},\,\mathbf{k}}^{(1)}+
\int_{-\infty}^\infty \frac{{\cal{V}}d^3\mathbf{q}'}{(2\pi)^3}\sum\limits_{\mathbf{e}^{(\mathbf{q}')}}
\frac{V_{\mathbf{k}-\mathbf{q},\,\mathbf{k}-\mathbf{q}'-\mathbf{q}}^{\left(0\right)}V_{\mathbf{k}-\mathbf{q}'-\mathbf{q},\,\mathbf{k}-\mathbf{q}'}^{\left(0\right)}V_{\mathbf{k}-\mathbf{q}',\,\mathbf{k}}^{\left(1\right)}}
{[\varepsilon_\mathbf{k}-\varepsilon_{\mathbf{k}-\mathbf{q}'-\mathbf{q}}+\hbar\omega- \hbar c(q'+q)+i0][\varepsilon_\mathbf{k}-\varepsilon_{\mathbf{k}-\mathbf{q}'}+\hbar\omega- \hbar cq'+i0]}\nonumber\\
&\approx V_{\mathbf{k}-\mathbf{q},\,\mathbf{k}}^{(1)}-i\pi\int_{-\infty}^\infty \frac{{\cal{V}}d^3\mathbf{q}'}{(2\pi)^3}\sum\limits_{\mathbf{e}^{(\mathbf{q}')}}
\frac{V_{\mathbf{k}-\mathbf{q},\,\mathbf{k}-\mathbf{q}'-\mathbf{q}}^{\left(0\right)}V_{\mathbf{k}-\mathbf{q}'-\mathbf{q},\,\mathbf{k}-\mathbf{q}'}^{\left(0\right)}V_{\mathbf{k}-\mathbf{q}',\,\mathbf{k}}^{\left(1\right)}}
{\varepsilon_\mathbf{k}-\varepsilon_{\mathbf{k}-\mathbf{q}'-\mathbf{q}}+\hbar\omega- \hbar c(q'+q)}\delta(\varepsilon_\mathbf{k}-\varepsilon_{\mathbf{k}-\mathbf{q}'}+\hbar\omega- \hbar cq')\approx\nonumber\\
&\approx V_{\mathbf{k}-\mathbf{q},\,\mathbf{k}}^{(1)}+i\pi\int_{-\infty}^\infty \frac{{\cal{V}}d^3\mathbf{q}'\delta(q'-\omega/c)}{(2\pi)^3(\hbar c)^2q'}\sum\limits_{\mathbf{e}^{(\mathbf{q}')}}
{V_{\mathbf{k}-\mathbf{q},\,\mathbf{k}-\mathbf{q}'-\mathbf{q}}^{\left(0\right)}V_{\mathbf{k}-\mathbf{q}'-\mathbf{q},\,\mathbf{k}-\mathbf{q}'}^{\left(0\right)}V_{\mathbf{k}-\mathbf{q}',\,\mathbf{k}}^{\left(1\right)}}
\nonumber\\
&\approx e\sqrt{\frac{2\pi\hbar }{cq\cal{V}}}
\left[\left(\mathbf{v_k}\cdot\mathbf{e}^{(\mathbf{q})}\right)J_{1}(\xi_\mathbf{q}^-)e^{i\varphi_\mathbf{q}}-
v_0\frac{{e}^{(\mathbf{q})}_x+i{e}^{(\mathbf{q})}_y}{2}
\right]+
\sqrt{\frac{2\pi\hbar }{cq\cal{V}}}\left({e\pi^2}\right)\left(\frac{v_0}{c}\right)^2\left(\frac{e^2}{\hbar c}\right)\left[{e_y^{(\mathbf{q})}}\cos\varphi_\mathbf{q}-
{e_x^{(\mathbf{q})}}\sin\varphi_\mathbf{q}\right]\nonumber\\
&\times J_{1}(\xi_{\mathbf{q}}^-)\int_{-\infty}^\infty \frac{d^3\mathbf{q}'\delta(q'-\omega/c)}{(2\pi)^3{q'}^2}\sum\limits_{\mathbf{e}^{(\mathbf{q}')}}
\left[v_{\mathbf{k}x}\left|e_x^{(\mathbf{q}')}\right|^2+
iv_{\mathbf{k}y}\left|e_y^{(\mathbf{q}')}\right|^2\right]\nonumber\\
&\approx -\sqrt{\frac{2\pi\hbar }{cq\cal{V}}}\left(\frac{ev_0}{2}\right)\left[
{{e}^{(\mathbf{q})}_x+i{e}^{(\mathbf{q})}_y}+\frac{\left(\mathbf{v_k}\cdot
\mathbf{e}^{(\mathbf{q})}\right)}{c}\left({n}^{(\mathbf{q})}_x+i{n}^{(\mathbf{q})}_y\right)\right]-\sqrt{\frac{2\pi\hbar }{cq\cal{V}}}\left(\frac{e\pi^2}{2}\right)\left(\frac{v_0}{c}\right)^3\left(\frac{e^2}{\hbar c}\right)\left[{e_y^{(\mathbf{q})}}n_x^{(\mathbf{q})}-
{e_x^{(\mathbf{q})}}n_y^{(\mathbf{q})}\right]\nonumber\\
&\times\int_{-\infty}^\infty \frac{d^3\mathbf{q}'\delta(q'-\omega/c)}{(2\pi)^3{q'}^2}\sum\limits_{\mathbf{e}^{(\mathbf{q}')}}
\left[v_{\mathbf{k}x}\left|e_x^{(\mathbf{q}')}\right|^2+
iv_{\mathbf{k}y}\left|e_y^{(\mathbf{q}')}\right|^2\right]\nonumber\\
&=-\sqrt{\frac{2\pi\hbar }{cq\cal{V}}}\left(\frac{ev_0}{2}\right)\left[
{{e}^{(\mathbf{q})}_x+i{e}^{(\mathbf{q})}_y}+\frac{\left(\mathbf{v_k}\cdot
\mathbf{e}^{(\mathbf{q})}\right)}{c}\left({n}^{(\mathbf{q})}_x+i{n}^{(\mathbf{q})}_y\right)\right]-\sqrt{\frac{2\pi\hbar }{cq\cal{V}}}\left(\frac{e\pi^2}{2}\right)\left(\frac{v_0}{c}\right)^3\left(\frac{e^2}{\hbar c}\right)\left[{e_y^{(\mathbf{q})}} n_x^{(\mathbf{q})}-
{e_x^{(\mathbf{q})}} n_y^{(\mathbf{q})}\right]\nonumber\\
&\times\int_0^{\pi}\frac{d\theta_{\mathbf{q}'}\sin\theta_{\mathbf{q}'}}{(2\pi)^3}
\int_0^{2\pi}d\varphi_{\mathbf{q}'}
[v_{\mathbf{k}x}(1-\sin^2\theta_{\mathbf{q}'}\cos^2\varphi_{\mathbf{q}'})+iv_{\mathbf{k}y}(1-\sin^2\theta_{\mathbf{q}'}\sin^2\varphi_{\mathbf{q}'})]\nonumber\\
&=
-{ev_0}\sqrt{\frac{\pi\hbar}{2cq\cal{V}}}\left[
{{e}^{(\mathbf{q})}_x+i{e}^{(\mathbf{q})}_y}+\frac{\left(\mathbf{v_k}\cdot
\mathbf{e}^{(\mathbf{q})}\right)}{c}\left({n}^{(\mathbf{q})}_x+i{n}^{(\mathbf{q})}_y\right)+\frac{1}{3}
\left(\frac{v_0}{c}\right)^2\left(\frac{e^2}{\hbar c}\right)\left({e_y^{(\mathbf{q})}} n_x^{(\mathbf{q})}-
{e_x^{(\mathbf{q})}} n_y^{(\mathbf{q})}\right)
\left(\frac{v_{\mathbf{k}x}}{c}+i\frac{v_{\mathbf{k}y}}{c}\right)\right].
\end{align}
Substituting the amplitude \eqref{T00} into Eq.~\eqref{wF3}, the probability of photon emission \eqref{wF3} takes its final form
\begin{equation}\label{W1}
W_\mathbf{k}(\mathbf{q})\approx W^{(a)}_\mathbf{k}(\mathbf{q})+{w}_\mathbf{k}(\mathbf{q}),
\end{equation}
where
\begin{align}\label{w1}
&{w}_\mathbf{k}(\mathbf{q})=\frac{2}{3}\left(\frac{\pi^2e^2}{\hbar q\omega\cal{V}}\right)\left(\frac{v_0}{c}\right)^4\left(\frac{e^2}{\hbar c}\right)
\left({v_{\mathbf{k}y}q_x}-{v_{\mathbf{k}x}q_y}
\right)\delta(q-q_1)\nonumber\\
&=\frac{2}{3}\left(\frac{\pi^2e^2v_0^2}{\hbar c\omega\cal{V}}\right)\left(\frac{v_0}{c}\right)^2\left(\frac{e^2}{\hbar c}\right)
\left(\frac{v_{\mathbf{k}y}}{c}\sin\theta_{\mathbf{q}}\cos\varphi_{\mathbf{q}}-\frac{v_{\mathbf{k}x}}{c}\sin\theta_{\mathbf{q}}\sin\varphi_{\mathbf{q}}
\right)\delta(q-q_1)
\end{align}
is the QED correction to the probability \eqref{w} arisen from the one-loop process pictured in Fig.~1b.

\begin{figure}[!h]
\includegraphics[width=.5\linewidth]{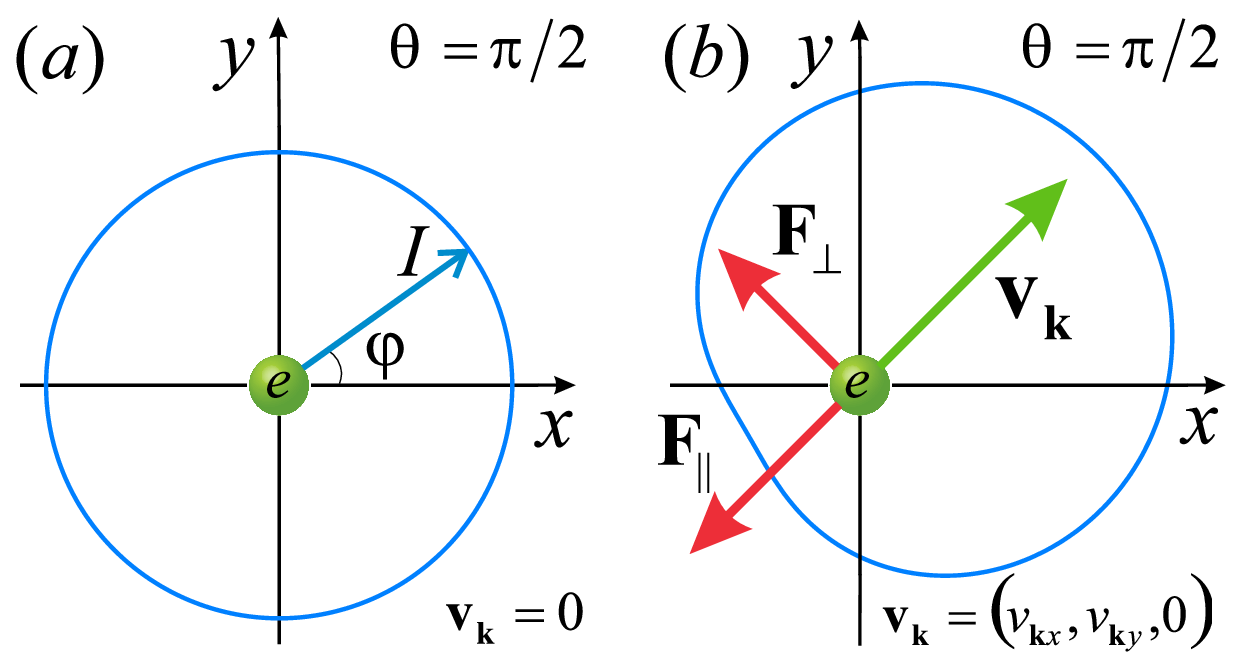}
\caption{The radiation pattern $I(\theta,\varphi)$ of an emitting electron $e$ in the $x,y$ plane: (a) for the electron at rest; (b) for the electron making forward movement with the velocity $\mathbf{v_k}$, where $\mathbf{F}_\|$ and $\mathbf{F}_\perp$ is the classical and quantum recoil force, respectively, acting on the electron due to the asymmetrical shape of the pattern.}
\end{figure}
Taking into account Eqs.~\eqref{W1}--\eqref{w1}, the total power emitted by an electron reads
\begin{equation}\label{I}
{{P}}=\sum_{\mathbf{q}}\hbar cqW_\mathbf{k}({\mathbf{q}})=\frac{\hbar c\cal{V}}{(2\pi)^3}\int_0^{2\pi}d\varphi_{\mathbf{q}}
\int_0^{\pi}d\theta_{\mathbf{q}}\sin\theta_{\mathbf{q}}\int_0^\infty dq \,q^3W_\mathbf{k}({\mathbf{q}})=
\int_0^{2\pi}d\varphi
\int_0^{\pi}d\theta\,I(\theta,\varphi)\sin\theta,
\end{equation}
where
\begin{equation}\label{I1}
I(\theta,\varphi)=\left.\frac{\hbar c\cal{V}}{(2\pi)^3}\int_0^\infty dq \,q^3W_\mathbf{k}({\mathbf{q}})\right|_{
\theta_{\mathbf{q}}=
\theta,\,
\varphi_{\mathbf{q}}=\varphi}
\end{equation}
is the angle distribution of the radiation emitted by the electron (the radiation pattern), which can be written explicitly as
\begin{align}\label{I0}
&I(\theta,\varphi)=\frac{e^2\omega^2v^2_0}{8\pi c^3}\Bigg[1+
\cos^2\theta+\frac{v_{\mathbf{k}x}}{c}\left(5\cos^2\theta+3\right)\sin\theta\cos\varphi+
\frac{v_{\mathbf{k}y}}{c}\left(5\cos^2\theta+3\right)\sin\theta\sin\varphi\nonumber\\
&+\frac{v_{\mathbf{k}z}}{c}\left(5\cos^2\theta+1\right)\cos{\theta}+\frac{2}{3}
\left(\frac{v_0}{c}\right)^2\left(\frac{e^2}{\hbar c}\right)
\left(\frac{v_{\mathbf{k}y}}{c}\sin{\theta}\cos{\varphi}-
\frac{v_{\mathbf{k}x}}{c}\sin{\theta}\sin{\varphi}\right)\Bigg].
\end{align}
This pattern is of symmetric shape for an electron at rest (see Fig.~2a) but it acquires asymmetry if the electron moves (see Fig.~2b). Due to such an asymmetric photon emission, the resultant momentum transferred to an emitting electron per unit time (the recoil force) differs from zero and reads
\begin{equation}\label{F1}
\mathbf{F}=-\sum_{\mathbf{q}}\hbar\mathbf{q}W_{\mathbf{k}}({\mathbf{q}})=-
\frac{1}{c}\int_0^{2\pi}d\varphi
\int_0^{\pi}d\theta\,\mathbf{n}I(\theta,\varphi)\sin\theta =\mathbf{F}_\|+\mathbf{F}_\perp,
\end{equation}
where $\mathbf{n}=\mathbf{r}/r=(\sin\theta\cos\varphi,\,
\sin\theta\sin\varphi,\,\cos\theta)$
is the unit radius vector, $\mathbf{F}_\|$ is the normal classical recoil force \eqref{LAD} directed oppositely to the electron velocity $\mathbf{v_k}$,
\begin{equation}\label{FK}
\mathbf{F}_\perp=-\sum_{\mathbf{q}}\hbar\mathbf{q}w_\mathbf{k}({\mathbf{q}})=\frac{1}{9}\left(\frac{e^4E^2_0}{m^2_ec^5}\right)
\left(\frac{v_0}{c}\right)^2\left(\frac{e^2}{\hbar c}\right)[\mathbf{L}\times\mathbf{v}_\mathbf{k}]
\end{equation}
is the anomalous recoil force directed perpendicularly to the velocity $\mathbf{v_k}$, and $\mathbf{L}$ is the unit vector along the field angular momentum, which defines the field polarization: The vector $\mathbf{L}=(0,\,0,\,1)$ describes the clockwisely polarized field \eqref{A0}, whereas the vector $\mathbf{L}=(0,\,0,\,-1)$ corresponds to the counterclockwise field polarization.

It should be noted that the emission processes described by the loop and loopless Feynman diagrams make substantially different contributions to the radiation reaction~\eqref{F1}. Mathematically, this follows from the fact that the anomalous recoil force~\eqref{FK} appears due to the singularities of the probability amplitude~\eqref{T0}, which arise from virtual photons [see the comments on Eqs.~\eqref{T0} and \eqref{T1} above]. In contrast to the loop diagrams, the loopless ones do not involve virtual photons and, therefore, their amplitudes are devoid of the singularities. As a consequence, the two emission processes pictured in Fig.~1 describe the radiation reaction~\eqref{F1} accurately within the leading order of electron-photon interaction: The one-vertex process makes main contribution to the normal recoil force~\eqref{LAD}, whereas the one-loop process is responsible for the anomalous force~\eqref{F1}. It should be stressed that the one-vertex photon emission pictured in Fig.~1a is the direct emission process which does not involve intermediate electron-photon states and, therefore, is similar physically to the classical process of electromagnetic emission. As a consequence, it results in the classical recoil force~\eqref{LAD}, where the Planck constant, $\hbar$, is cancelled. As to the one-loop photon emission pictured in Fig.~1b, it is the indirect process going through the intermediate state with the virtual photon $\mathbf{q}'$. It has no analogue within classical electrodynamics and, therefore, yields the anomalous recoil force~\eqref{FK} involving the fine structure constant, $\alpha=e^2/\hbar c$, which is of purely quantum nature.

\begin{figure}[!h]
\includegraphics[width=.5\linewidth]{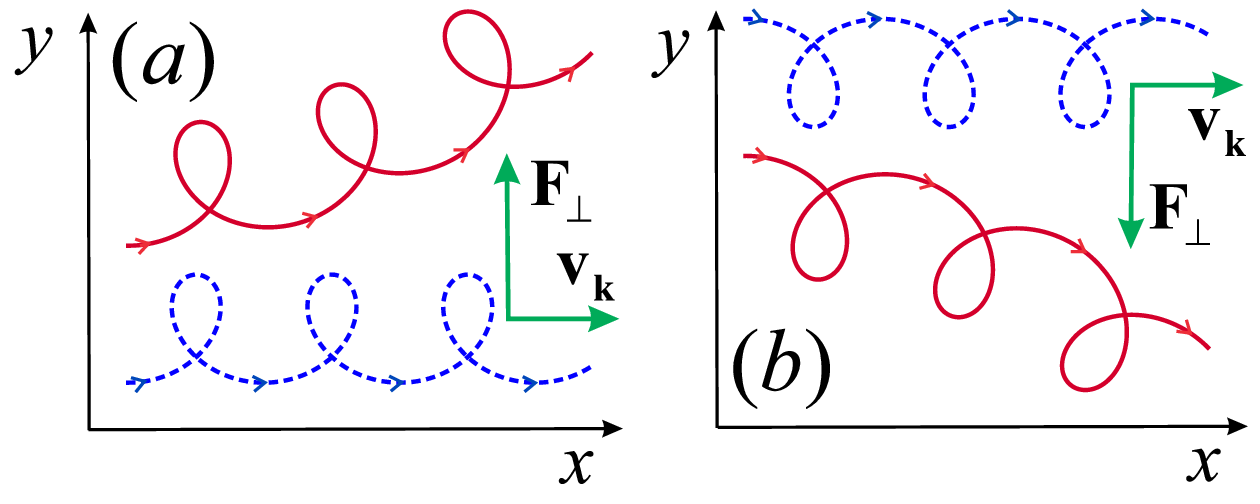}
\caption{Electron trajectories under a circularly polarized field rotating in the $x,y$ plane: (a) for clockwise field polarization; (b) for counterclockwise field polarization. The trajectories describing the electron motion within the classical electrodynamics are plotted by the dashed lines, whereas the solid lines indicate these trajectories under the quantum recoil force, $\mathbf{F}_\perp$, directed perpendicularly to the forward movement velocity, $\mathbf{v_k}$.}
\end{figure}
Evolution of the electron velocity $\mathbf{v_k}$  under the force \eqref{F1} can be described by the Newton equation,
\begin{equation}\label{NE}
m_e\dot{\mathbf{v}}_\mathbf{k}=-\frac{2}{3}\left(\frac{e^4E^2_0}{m^2_ec^5}\right)\mathbf{v}_\mathbf{k}+\frac{1}{9}\left(\frac{e^4E^2_0}{m^2_ec^5}\right)
\left(\frac{v_0}{c}\right)^2\left(\frac{e^2}{\hbar c}\right)[\mathbf{L}\times\mathbf{v}_\mathbf{k}],
\end{equation}
which leads to the electron trajectory pictured schematically in Fig.~3. Since the quantum recoil force (the second term in the right-hand side of this equation) is directed perpendicularly to the electron velocity, it does not decelerate an electron in contrast to the classical recoil force (the first term there) but bends its trajectory similarly to the Lorentz force acting on an electron in a magnetic field. Such an analogy has the deep physical nature since both a high-frequency circularly polarized field and a stationary magnetic field break the time-reversal symmetry. As a consequence, a circularly polarized field can induce various magnetic-like effects, including the spin polarization~\cite{Kibis_2022}, bandgap opening in solids~\cite{Kibis_2010,Wang_2013,Sie_2015,Cavalleri_2020}, persistent currents in conductors~\cite{Kibis_2011,Melnikov_2021}, etc. It should be stressed that the spatial asymmetry of photon emission responsible for the quantum recoil force \eqref{FK} --- which is described by the last term in Eq.~\eqref{I0} --- also arises from the broken time-reversal symmetry. Therefore, the discussed anomalous radiation reaction acting on a charged particle perpendicularly to its velocity takes place in any electromagnetic field with nonzero angular momentum $\mathbf{L}$ (circularly and elliptically polarized fields) and vanishes only in the case of linearly polarized field with $\mathbf{L}=0$. It follows from the aforesaid, particularly, that Eq.~\eqref{FK} should be considered as a quantum correction to the Landau-Lifshitz equation~\cite{Landau_2} describing the classical radiation reaction force in an electromagnetic field of most general form. It should be noted also that the angular momentum of a circularly polarized field, $\mathbf{L}$,  is codirectional to the angular momentum of a charge rotating under the field. This means that the quantum recoil force \eqref{FK} is similar phenomenologically to the classical Magnus force, $\mathbf{F}_M\propto[\mathbf{L}\times\mathbf{v}]$, acting on a body which rotates with the angular momentum $\mathbf{L}$ and makes forward movement with the velocity $\mathbf{v}$ in a medium with friction (see, e.g., Ref.~\onlinecite{Clancy_book}). Treating the classical recoil force \eqref{LAD} as a radiation friction, the quantum recoil force \eqref{FK} can be considered qualitatively as a specific kind of the Magnus force arisen from this friction acting on a rotating charge.

Although the present theory is developed for the homogeneous field \eqref{A0}, the force~\eqref{FK} very weakly depends on spatial profile of the field if its characteristic scale much exceeds the radius of circular electron trajectory, $r_0=|v_0|/\omega$. As to the time profile of the field (the laser pulse duration), its characteristic scale should much exceed the radiative life time \eqref{tau}. Since the radiative life time, $\tau$, must be large enough to satisfy the condition $\omega\tau\gg1$, it is most suitable to detect the force~\eqref{FK} with non-relativistic laser fields which can be easily generated in the continuous-wave regime.
Discussing dependence of the force~\eqref{FK} on the field strength, $E_0$, it should be noted that the Floquet functions \eqref{psi} are exact solutions of the non-relativistic Schr\"odinger problem for an electron in the circularly polarized background field \eqref{A0} of any strength. Therefore, the developed theory is applicable to the background field of the strength $E_0$ satisfying only the non-relativistic condition, $v_0/c=eE_0/m_e\omega c\ll1$. To keep the ratio $v_0/c$ to be small enough within applicability of the developed non-relativistic theory, one can consider laser fields of the strength $E_0$ up to $10^{11}$~V/m. Estimating the force \eqref{FK} quantitatively, it should be noted that the electric field amplitude $E_0$ can reach $10^{10}$~V/m for an electromagnetic wave generated by an ordinary hundred-kilowatt laser with the wavelengths $\lambda\sim10^{-6}$~m. In such a field, the quantum force \eqref{FK} yields the electron acceleration $a_\perp=F_\perp/m_e\sim10^3(v_\mathbf{k}/c)$~m/s$^2$, which can be macroscopically large even for non-relativistic electrons satisfying the condition $v_\mathbf{k}\ll c$. Thus, the discussed anomalous radiation reaction is the rare case of macroscopic QED effect which can be observable for slow electrons in laser fields of non-relativistic intensity.

To extend the present theory over the relativistic case, one needs to start from the relativistic Hamiltonian describing the considered emission problem,
$\hat{\cal H}=\hat{\cal H}_D+\hat{V}^{(\pm\mathbf{q})}$, where $\hat{\cal H}_D=c\boldsymbol{\alpha}(\hat{\mathbf{p}}-e\mathbf{A}/c)+\beta m_ec^2$ is the Dirac Hamiltonian for an electron in the background field \eqref{A0}, and $\boldsymbol{\alpha},\beta$ are the Dirac matrices~\cite{Landau_4}. Since the eigenfunctions (bispinors) of the Dirac Hamiltonian, $\Psi_{\mathbf{k}}$, form the complete orthonormal function system, the solution of the Dirac problem with the Hamiltonian $\hat{\cal H}$ can be sought as an expansion $\Psi=\sum_{\mathbf{k}'}c_{\mathbf{k}'}(t)\Psi_{\mathbf{k}'}$.
Substituting this into the Dirac equation, $\hat{\cal H}\Psi=i\hbar\partial_t\Psi$, we arrive at the Dirac equation written in the Floquet representation,
\begin{equation}\label{dck1}
i\hbar\frac{dc_{\mathbf{k}'}(t)}{dt}=\sum_{\mathbf{k}}\langle\Psi_{\mathbf{k}'}|\hat{V}^{(\pm\mathbf{q})}|\Psi_{\mathbf{k}}\rangle c_{\mathbf{k}}(t),
\end{equation}
which defines the expansion coefficients $c_{\mathbf{k}}(t)$. Within the conventional relativistic quantum electrodynamics~\cite{Landau_4}, the matrix element of the electron interaction with a single photon reads
\begin{equation}\label{V01}
\langle\Psi_{\mathbf{k}'}|\hat{V}^{(\pm\mathbf{q})}|\Psi_{\mathbf{k}}\rangle=\frac{e}{c}
\int_{\cal{V}}\left(\mathbf{j}_{\mathbf{k}'\mathbf{k}}\cdot\mathbf{A}^{(\pm\mathbf{q})}\right)d^3\mathbf{r},
\end{equation}
where $\mathbf{A}^{(\pm\mathbf{q})}=\mathbf{e}^{(\mathbf{\pm q})}\sqrt{2\pi\hbar c/q\cal{V}}\,e^{\pm i(\mathbf{qr}-cqt)}$
is the photon wave function, and $\mathbf{j}_{\mathbf{k}'\mathbf{k}}=c\Psi^*_{\mathbf{k}'}\boldsymbol{\alpha}\Psi_{\mathbf{k}}$ is the relativistic transition current. Next, let us apply the expansion in powers of $1/c$, which is conventionally used to transform the relativistic Dirac equation into the non-relativistic Schr\"odinger equation (see, e.g., Sec.~33 in Ref.~\onlinecite{Landau_4}). Then the transition current reads~\cite{Landau_4}
\begin{equation}\label{jk1}
\mathbf{j}_{\mathbf{k}'\mathbf{k}}=c\Psi^*_{\mathbf{k}'}\boldsymbol{\alpha}\Psi_{\mathbf{k}}=\frac{1}{2m_e}(\psi^\ast_{\mathbf{k}'}\hat{\mathbf{p}}\psi_{\mathbf{k}}-
\psi_{\mathbf{k}}\hat{\mathbf{p}}\psi^\ast_{\mathbf{k}'})-\frac{e}{m_ec}\mathbf{A}
\psi^\ast_{\mathbf{k}'}\psi_{\mathbf{k}}+{\it O}(1/c^2)
\end{equation}
and the Dirac equation \eqref{dck1} turns into the Schr\"odinger equation \eqref{dck} as expected.
It follows from the aforesaid that the accurate transformation of the exact relativistic Dirac equation \eqref{dck1} into the approximate non-relativistic Schr\"odinger equation \eqref{dck} corresponds to neglecting the relativistically small terms only in the transition current \eqref{jk1}, whereas other $1/c$-dependent terms --- arisen in the electron-photon interaction from the photon wave function --- are not affected by this transformation [see, particularly, the $1/c$-dependent factor before the square bracket in Eq.~\eqref{Vmm}]. Mathematically, just these terms yield the factor $(v_0/c)^2$ in the quantum force \eqref{FK}. Therefore, Eq.~\eqref{FK} stays within applicability of the present non-relativistic theory despite the relativistic smallness of this factor. As to analysis of fast electrons moving in ultra-strong fields, it needs the accurate solving of the Dirac problem \eqref{dck1}, which goes beyond the scope of the present paper and should be done elsewhere.

\section{Conclusion}

It is shown theoretically that the emission of photons by an electron rotating under a high-frequency circularly polarized electromagnetic field yields the quantum recoil force acting on the electron perpendicularly to the velocity of its forward movement. In contrast to the classical Lorentz-Abraham-Dirac recoil force directed oppositely to the velocity, it does not decelerate the electron but substantially modifies its trajectory. Such an anomalous radiation reaction arises from the one-loop QED correction to the photon emission and can manifest itself in strong laser-generated fields.}

\end{document}